\documentclass[lettersize,journal]{IEEEtran}
\usepackage{amsmath,amsfonts}
\usepackage{algorithmic}
\usepackage{algorithm}
\usepackage{array}
\usepackage[caption=false,font=footnotesize,labelfont=rm,textfont=rm]{subfig}
\usepackage{textcomp}
\usepackage{stfloats}
\usepackage{url}
\usepackage{verbatim}
\usepackage{graphicx}
\usepackage{cite}
\usepackage{color}
\usepackage{bm}
\usepackage{threeparttable}
\usepackage{makecell}
\usepackage{multirow}

\begin{document}

\title{MagLive: Robust Voice Liveness Detection on Smartphones Using Magnetic Pattern Changes}

\author{Xiping Sun, Jing Chen, Cong Wu, Kun He, Haozhe Xu, Yebo Feng, Ruiying Du, Xianhao Chen

\thanks{
Xiping Sun, Kun He and Haozhe Xu are with Key Laboratory of Aerospace Information Security and Trusted Computing, Ministry of Education, School of Cyber Science and Engineering, Wuhan University, Wuhan, 430072, China (e-mail: \{xiping, hekun, haozhexu\}@whu.edu.cn).

Jing Chen is with Key Laboratory of Aerospace Information Security and Trusted Computing, Ministry of Education, School of Cyber Science and Engineering, Wuhan University, Wuhan, 430072, China, and also with Rizhao Institute of Information Technology, Wuhan University, Rizhao, 276800, China (e-mail: chenjing@whu.edu.cn).

Cong Wu and Yebo Feng are with School of Computer Science and Engineering, Nanyang Technological University, Singapore (e-mail: \{cong.wu, yebo.feng\}@ntu.edu.sg).

Ruiying Du is with Key Laboratory of Aerospace Information Security and Trusted Computing, Ministry of Education, School of Cyber Science and Engineering, Wuhan University, Wuhan, 430072, China, and also with Collaborative Innovation Center of Geospatial Technology, Wuhan, 430079, China (e-mail: duraying@whu.edu.cn).

Xianhao Chen is with the Department of Electrical and Electronic Engineering and HKU Musketeers Foundation Institute of Data Science, University of Hong Kong, Pok Fu Lam, Hong Kong SAR, China (e-mail: xchen@eee.hku.hk).

}
}

\maketitle

\begin{abstract}
	Voice authentication has been widely used on smartphones.
	However, it remains vulnerable to spoofing attacks, where the attacker replays recorded voice samples from authentic humans using loudspeakers to bypass the voice authentication system.
	In this paper, we present MagLive, a robust voice liveness detection scheme designed for smartphones to mitigate such spoofing attacks.
	MagLive leverages the differences in magnetic pattern changes generated by different speakers (i.e., humans or loudspeakers) when speaking for liveness detection, which are captured by the built-in magnetometer on smartphones. 	
	To extract effective and robust magnetic features, MagLive utilizes a TF-CNN-SAF model as the feature extractor, which includes a time-frequency convolutional neural network (TF-CNN) combined with a self-attention-based fusion (SAF) model.
	Supervised contrastive learning is then employed to achieve user-irrelevance, device-irrelevance, and content-irrelevance.
	MagLive imposes no additional burden on users and does not rely on active sensing or specialized hardware.
	We conducted comprehensive experiments with various settings to evaluate the security and robustness of MagLive. 
	Our results demonstrate that MagLive effectively distinguishes between humans and attackers (i.e., loudspeakers), achieving an average balanced accuracy (BAC) of 99.01\% and an equal error rate (EER) of 0.77\%.
\end{abstract}


\section{Introduction}
\label{sec:introduction}

\IEEEPARstart{V}{oice} authentication technologies, as a well-known form of biometrics, have seen a marked increase in adoption for executing sensitive operations on modern smartphones.
These operations encompass a range of activities, from secure logins to mobile banking transactions. 
Notable implementations include WeChat's Voiceprint feature, 
which facilitates user login through voice passwords~\cite{wechat}, 
and Citi's deployment of voice biometrics for customer identification~\cite{citi}. 
Despite the convenience and user-friendly nature of these systems, the inherent public exposure of human voices introduces significant vulnerabilities. 
Given the easily accessible nature of voice data, these systems are particularly prone to spoofing attacks wherein attackers could record and manipulate voice samples, then replay them with loudspeakers to unauthorizedly access secure services.
Such vulnerabilities not only compromise personal privacy but also pose substantial risks of financial loss and unauthorized access to sensitive information.

\begin{figure}[!t]
	\centering
	\subfloat[Scenario]{\includegraphics[width = 0.45\linewidth]{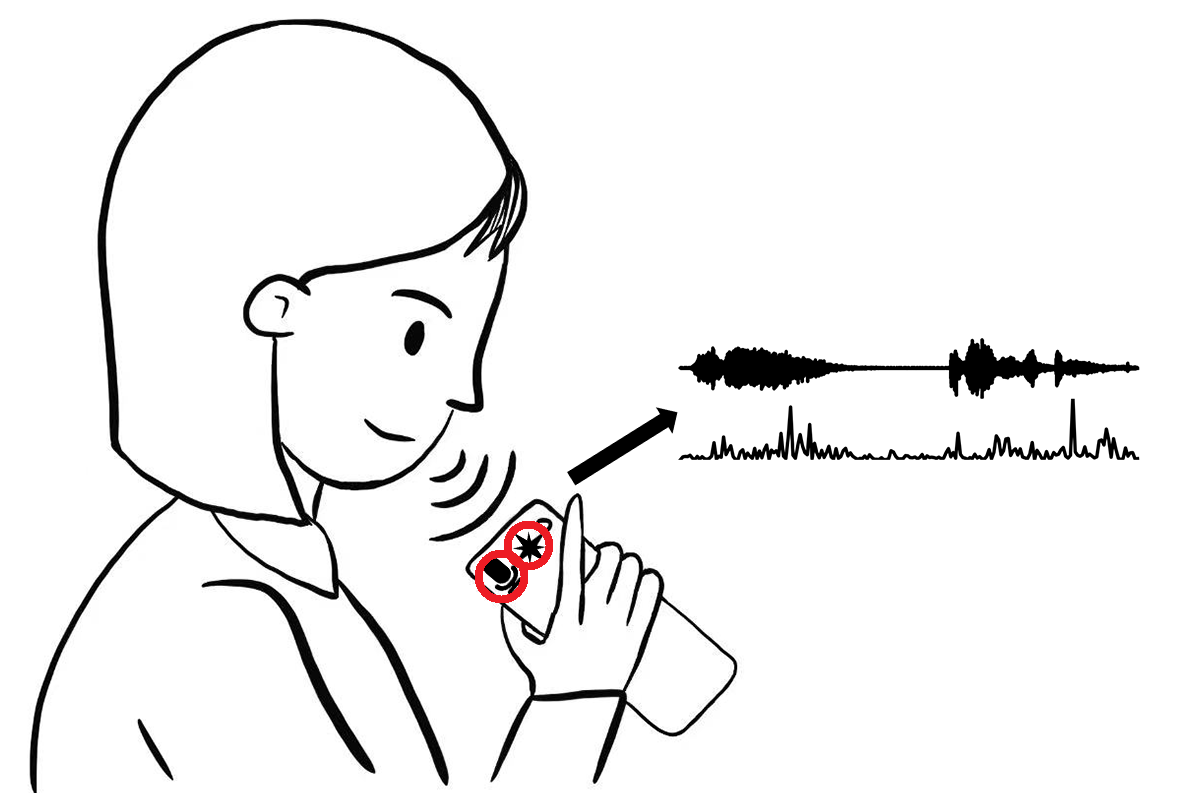}
		\label{fig:user}}
	\subfloat[Liveness detection]{\includegraphics[width = 0.5\linewidth]{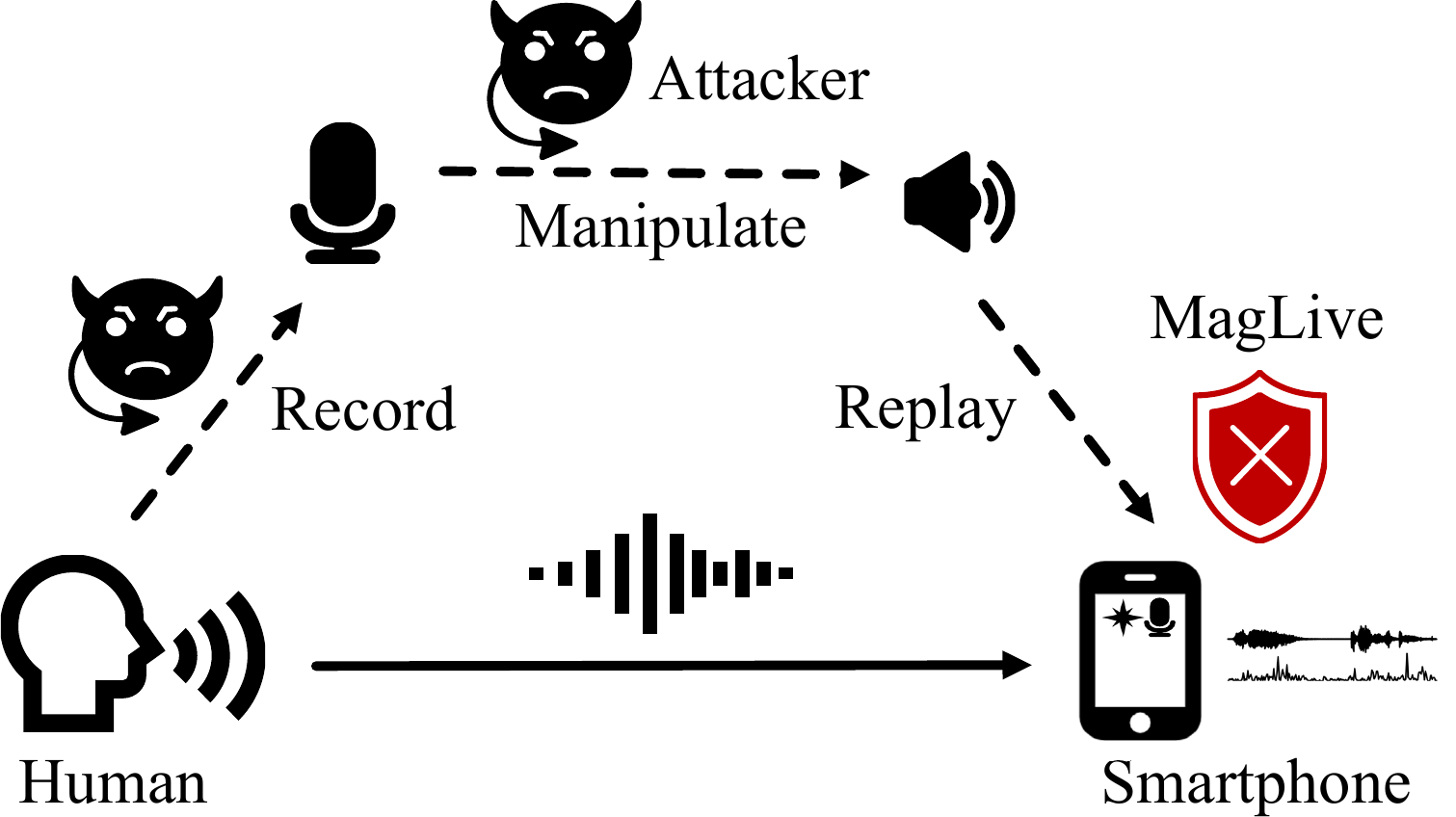}
		\label{fig:attack}}
	\caption{
		Illustration of MagLive.
		(a) It uses the built-in magnetometer and microphone on the smartphone for voice liveness detection.
		(b) It detects the liveness of the voice to determine whether it is from an authentic human or artificially reproduced.
	}
	\label{fig:maglive_show}
\end{figure}

Efforts to counteract spoofing attacks have led to advancements in liveness detection technology, designed to determine whether a voice is genuinely human or artificially reproduced and replayed by a loudspeaker.
Research in this area spans various devices like smartphones, smart speakers, and wearables. 
Techniques include using microphone arrays~\cite{li2021robust,meng2022your,yang2023voshield,lee2020using} and other specialized hardware (wireless, mmWave radars)~\cite{meng2018wivo,zhao2021anti,pradhan2019combating,li2020vocalprint} for smart speakers, and methods tailored for wearable devices~\cite{shi2020wearid,feng2017continuous,blue20182ma}.

Voice liveness detection in smartphones can be categorized into three key areas, each exploiting distinct features of voice interaction. 
Firstly, sound field features analyze the acoustic energy created as the sound propagates over the air~\cite{yan2019catcher}, but require the user to remain stationary gesture for accurate results.
Secondly, human features target the biological mechanisms of voice production~\cite{zhang2017hearing,lu2018lippass,wu2019lvid,chen2021chestlive,wang2019secure,zhang2016voicelive}, yet may introduce discomfort with the need for high-frequency sound active sensing or reliance on specialized hardware. 
Thirdly, loudspeaker features focus on detecting anomalies in speaker output~\cite{blue2018hello,ahmed2020void,chen2017you}. 
However, some struggle against advanced attacks~\cite{wang2020differences} that closely imitate human voices~\cite{blue2018hello,ahmed2020void}, while others are unsuitable for diverse environments and demand specific user actions, compromising practicality and convenience~\cite{chen2017you}.

\textbf{MagLive}.
In this work, we introduce MagLive, a robust approach for detecting the liveness of voices on smartphones by utilizing changes in magnetic patterns, overcoming the shortcomings of existing methods that lack user-friendliness and security.
The key idea is to discern the distinctive variations in magnetic pattern generated by human speech as opposed to those produced by loudspeakers, as depicted in Fig.~\ref{fig:maglive_show}.
This technique utilizes the inherent magnetometer and microphone already equipped in smartphones, enabling voice authenticity detection without the need for active signal sensing or specialized hardware.
MagLive features minimal operational constraints and maintains its effectiveness across diverse environmental conditions, setting a new standard in voice liveness detection technology.

The development of MagLive presents three primary challenges. 
The first challenge is that the magnetic field changes induced by speakers are minute and weak, causing useful signals to be potentially submerged by noise in the magnetometer. 
We tackle this by first filtering out noise and neutralizing the effects of the Earth's magnetic field. 
Then, we leverage the voice data to aid in the segmentation of the magnetometer data.
The second challenge is the intricate nature of magnetic field variations caused by speakers, which complicates the extraction of relevant magnetic patterns for voice detection.
To overcome this, we deploy a TF-CNN-SAF model to achieve precise feature representation, combining a time-frequency convolutional neural network (TF-CNN) with a self-attention-based fusion (SAF) model.
The third challenge stems from the susceptibility of magnetometer data to diverse external factors, including user interactions, device differences, voice variations, and environmental noise.
We employ a supervised contrastive learning approach to maximize the differences between samples from humans and loudspeakers, while minimizing the differences between samples within the same class.

\begin{figure}[!t]
	\centering
	\includegraphics[width = 1\linewidth]{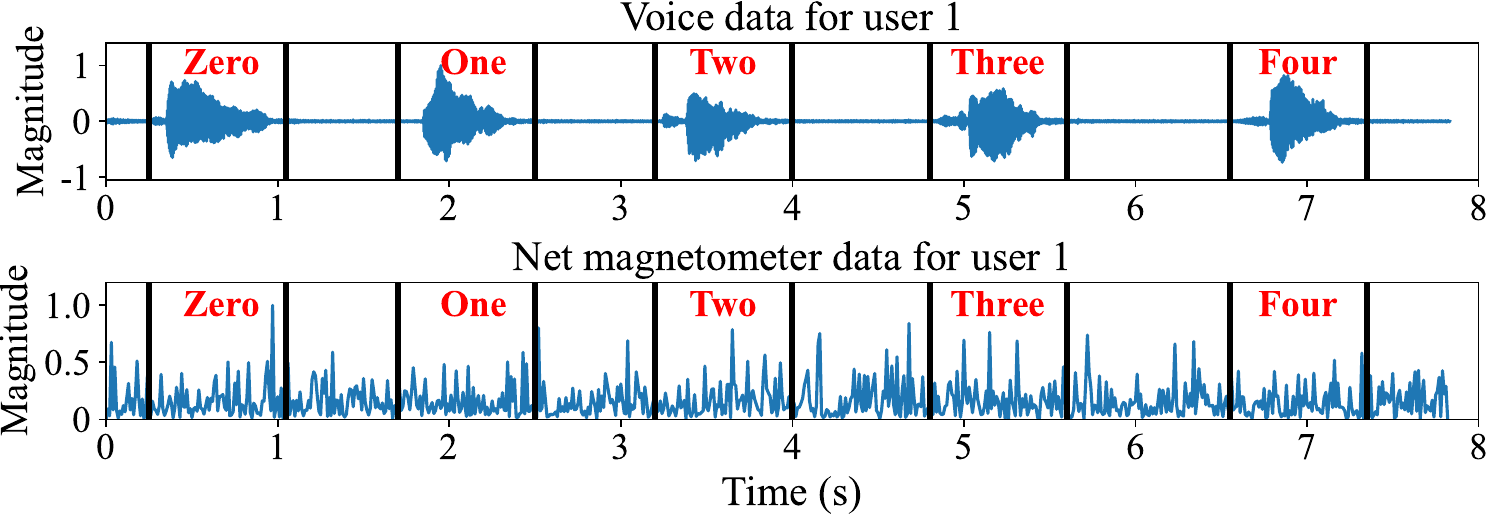}
	\caption{An example of user 1 speaking digits from zero to four (human).}
	\label{fig:human1}
\end{figure}

\begin{figure}[!t]
	\centering
	\includegraphics[width = 1\linewidth]{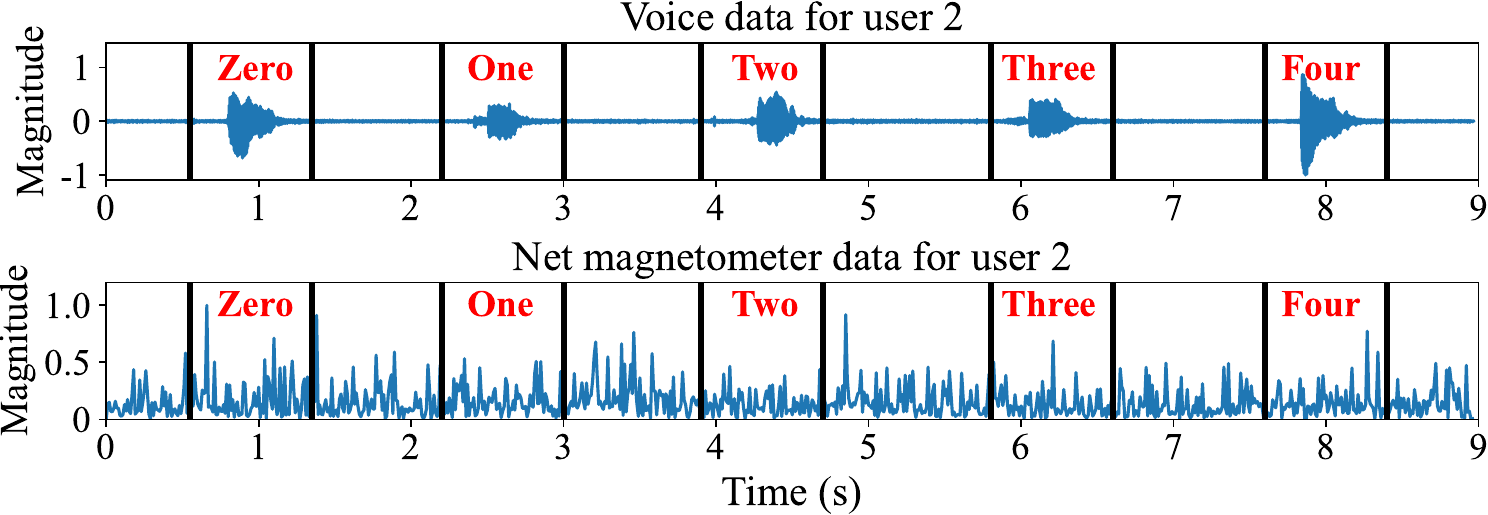}
	\caption{An example of user 2 speaking digits from zero to four (human).}
	\label{fig:human2}
\end{figure}

\textbf{Novelty}.
Significantly different from previous work, especially that of Chen \emph{et al.}~\cite{chen2017you}, the novelty of our paper lies in the following aspects:
1) Effective feature extraction: MagLive is the first work to explore the effective changes in magnetic patterns associated with speaking, rather than just the absolute value and changing rate of the magnetic field.
2) Suitability for diverse environments: MagLive is designed to maintain its effectiveness across diverse environmental conditions, instead of simply relying on magnetic strength and rate of change thresholds, which are less adaptable to various environments.
3) Little usage constraints: MagLive has minimal operational constraints. It does not require active sensing or the user to move the smartphone along a predefined trajectory.

Our contributions can be summarized as follows:
\begin{itemize}
	\item	
	To the best of our knowledge, MagLive is the first work to explore magnetic pattern changes associated with speaking for robust voice liveness detection, providing a defense against spoofing attacks on smartphones.
	\item	
	We devise a series of preprocessing methods to isolate the segments of magnetometer data induced by speech.
	We design 
	1) a feature extraction method based on the TF-CNN-SAF model to derive effective and robust features from magnetometer data, and 
	2) a supervised contrastive learning-based training method to ensure user-irrelevance, device-irrelevance, and content-irrelevance.
	\item
	We conduct comprehensive experiments across real devices and various settings to evaluate the performance of MagLive, achieving an average balanced accuracy (BAC) of 99.01\% and an equal error rate (EER) of 0.77\%.
\end{itemize}

\begin{figure}[!t]
	\centering
	\includegraphics[width = 1\linewidth]{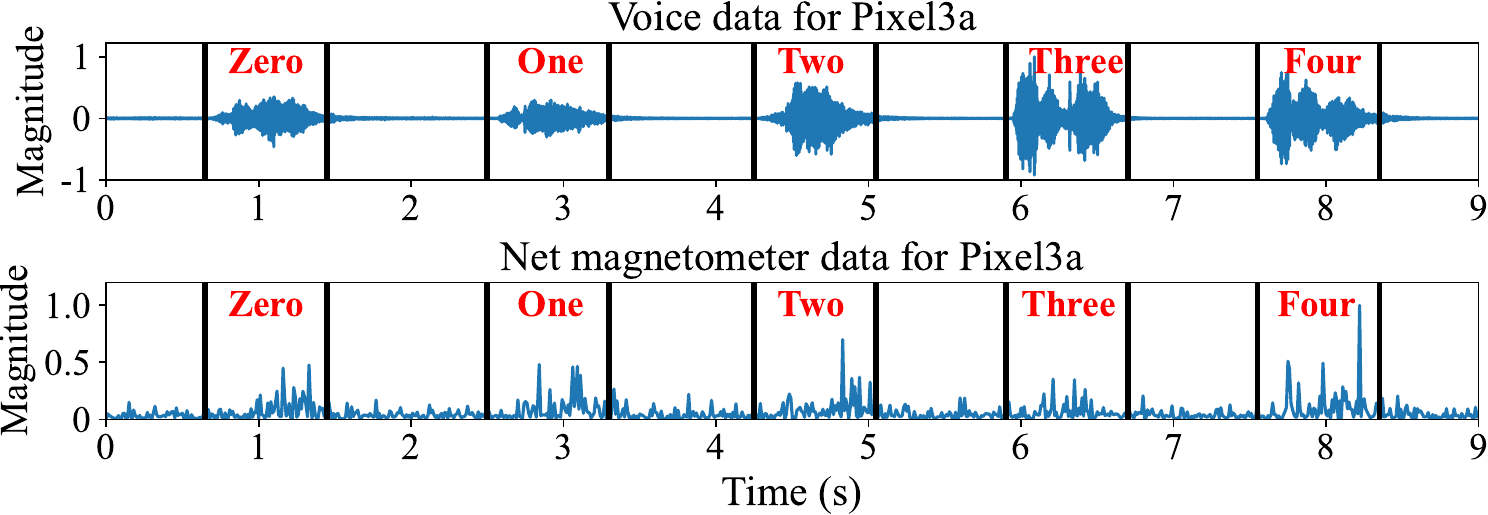}
	\caption{An example of Pixel3a replaying the speech of User 1 (loudspeaker).}
	\label{fig:pixel3a}
\end{figure}

\begin{figure}[!t]
	\centering
	\includegraphics[width = 1\linewidth]{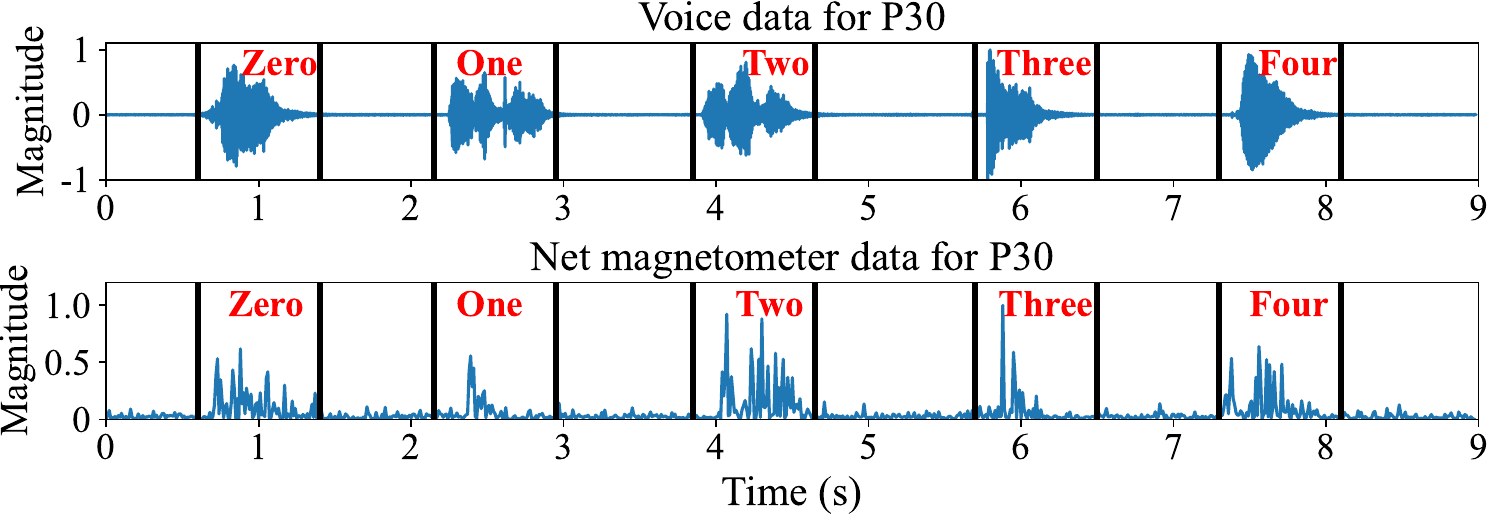}
	\caption{An example of P30 replaying the speech of user 1 (loudspeaker).}
	\label{fig:p30}
\end{figure}

\section{Preliminaries}
\label{sec:preliminaries}
In this section, we first introduce the magnetic effect of speakers, and then show the motivating examples.

\subsection{Magnetic Effect of Speakers}
\label{magnetic_effect}

The mechanisms behind sound production in humans and loudspeakers are distinct. 
Human speech is created through the coordination of vocal cords and various other organs, working in unison to produce sound~\cite{blue2022you}. 
On the other hand, an electric loudspeaker operates on electromagnetic induction~\cite{liu2023magbackdoor}, converting electrical currents into sound.
This process results in a specific magnetic signature unique to loudspeakers.

\begin{figure}[!t]
	\centering
	\includegraphics[width = 0.55\linewidth]{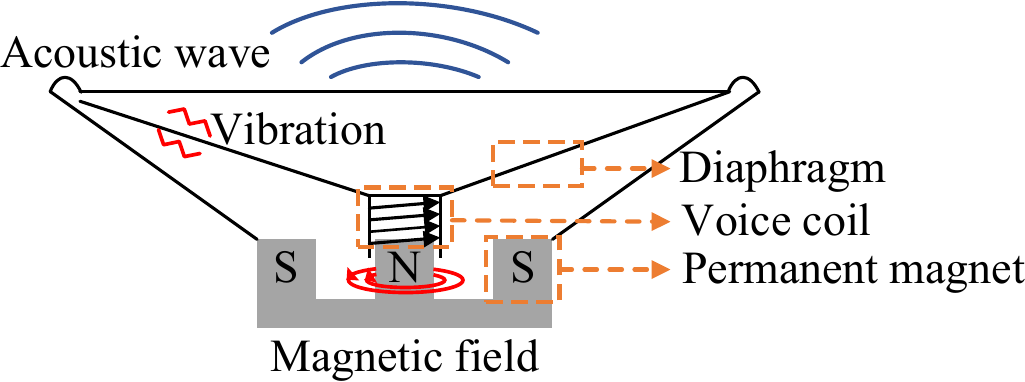}
	\caption{The mechanical structure of an electric loudspeaker.}
	\label{fig:loudspeaker}
\end{figure}

Fig.~\ref{fig:loudspeaker} illustrates the basic components of an electric loudspeaker: a permanent magnet, a voice coil, and a diaphragm. 
The permanent magnet creates a steady magnetic field. 
When the loudspeaker is in use, electrical current passes through the voice coil, turning it into a temporary electromagnet that generates changing magnetic fields. 
This interaction causes the voice coil to move towards or away from the permanent magnet. 
The diaphragm, attached to the voice coil, vibrates in response to these movements, pushing against the air to create sound waves. 
These sound waves result from the shifts in the magnetic field around the coil, meaning that sound production in a loudspeaker is directly linked to variations in magnetic patterns.

\begin{figure*}[!t]
	\centering
	\includegraphics[width = 1\linewidth]{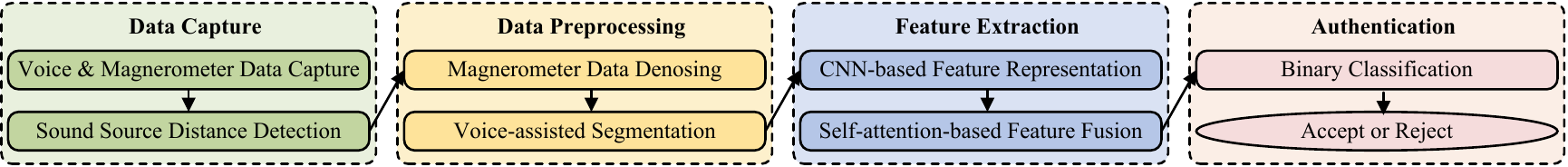}
	\caption{Workflow of MagLive.}
	\label{fig:workflow}
\end{figure*}

\subsection{Motivating Examples}

To investigate the variations in magnetic patterns associated with speech from different sources—namely, humans and loudspeakers—we carried out initial experiments. 
These involved recording both magnetic signals and voice data using the built-in magnetometer and microphone of a smartphone. 
For our experiments, we selected the iPhone 14 Pro as the authentication device and collected magnetometer data at its highest sampling rate of 100Hz.

%human
To investigate the magnetic signatures associated with human speech, we enlisted two volunteers to speak a sequence of numbers from zero to four in English.
The resulting voice and magnetometer data, normalized for comparison, are illustrated in Fig.\ref{fig:human1} for the first user and Fig.\ref{fig:human2} for the second. 
%loudspeaker
To assess the magnetic effects generated by loudspeakers, we played back the recording of the first user's speech through two different speaker models, Pixel3a and P30, without any direct contact with the testing smartphone. The outcomes of these tests are depicted in Fig.\ref{fig:pixel3a} and Fig.\ref{fig:p30}, showcasing the voice and magnetometer data during the simulated spoofing attacks.
%result
Distinct patterns can be observed from the magnetometer data triggered by loudspeakers compared to the natural human voice. 
These results are consistent with the analyses in Section~\ref{magnetic_effect}, which indicate that dynamic magnetic fields caused by loudspeakers when emitting sound create specific magnetic signature. 
Notably, the magnetic field changes varied not just between humans and loudspeakers but also among different individuals and spoofing devices, despite reproducing the same set of spoken digits. 
These variations underscore the unique magnetic footprints left by different sources and form the empirical basis for MagLive's design. 
Leveraging these distinct magnetic patterns, we refine our voice liveness detection methodology in MagLive.

\section{Overview of MagLive}
\label{sec:overview}
In this section, we first present the system overview of MagLive.
Then, we introduce the threat model and design goals.

\subsection{System Overview}
The basic idea of MagLive is that different speakers (i.e., humans or loudspeakers) generate unique magnetic pattern changes when speaking.
Therefore, MagLive utilizes the commercial smartphone's built-in microphone and magnetometer for magnetic sensing-based voice liveness detection.
Fig.~\ref{fig:workflow} shows the workflow of MagLive, which comprises four modules: data capture, data preprocessing, feature extraction, and authentication.

In our MagLive framework, the data capture module concurrently gathers voice and magnetometer data from the authentication device, typically a smartphone, with subsequent analysis to determine the distance between the sound source and the device. 
During data preprocessing, we refine the magnetometer data by eliminating noise and reducing the Earth's magnetic field's impact, using voice data to segment the magnetometer readings effectively.
The feature extraction module then isolates patterns of magnetic field fluctuations attributed to different sources, employing a TF-CNN-SAF model that combines a time-frequency convolutional neural network (TF-CNN) with a self-attention-based fusion (SAF) model for effective and robust feature derivation. 
This process is enhanced through supervised contrastive learning, tailored to ensure the system's independence from user identity, voice content, and device variation.
Finally, the authentication module processes these refined features through a binary classifier to ascertain the voice sample's authenticity and differentiate between human and non-human sources efficiently.

\subsection{Threat Model}
The goal of an attacker is to bypass the voice authentication system and perform sensitive operations.
State-of-the-art voice authentication systems are robust against human-based voice impersonation attacks, where an attacker tries to mimic a target user's voice timbre and prosody without machines~\cite{chen2017you}. 
Therefore, such attacks do not pose a real threat~\cite{kassis2023breaking}.
In this paper, we focus on a more realistic spoofing attack scenario, where attackers record and manipulate voice samples, then replay them using spoofing devices (primarily loudspeakers). 
Thus, considering the attacker’s ability and goal, spoofing attacks can be classified into three types~\cite{yan2022survey}:

\begin{itemize}
	\item \emph{Replay.} The attacker uses loudspeakers to replay recorded voice samples collected from the target user~\cite{pradhan2019combating}.
	\item \emph{Speech synthesis.} The attacker generates intelligible artificial voice that sounds like the target user from text and plays it via loudspeakers~\cite{wenger2021hello}.
	\item \emph{Voice conversion.} The attacker manipulates the voice of a human speaker so that it resembles the target user and plays it with loudspeakers~\cite{deng2023catch}.
\end{itemize}

Besides spoofing attacks, we also consider advanced attacks such as modulated attacks~\cite{wang2020differences} and adversarial attacks~\cite{yu2023smack}.
While the techniques of the above attacks differ, they rely on spoofing devices (primarily loudspeakers) to replay the manipulated voice samples, as shown in Fig.~\ref{fig:maglive_show}\subref{fig:attack}.
Therefore, we focus on detecting the distinct characteristics of loudspeakers to cope with these attacks.

\subsection{Design Goals}
A suitable liveness detection method for a voice authentication system should satisfy the following goals:
\begin{itemize}
	\item \emph{Security:} 
	it should effectively distinguish between legitimate samples and various spoofing samples, ensuring that only authentic users are granted access.
	\item \emph{Robustness:} 
	it should perform robustly across diverse conditions,
	such as different spoofing devices and authentication devices, various users, different voice contents, and different environmental settings.
	\item \emph{User-friendliness:} 
	it should be easy to use for users, requiring minimal user intervention and not relying on active sensing or specialized hardware.
\end{itemize}

\begin{figure}[!t]
	\centering
	\includegraphics[width = 1\linewidth]{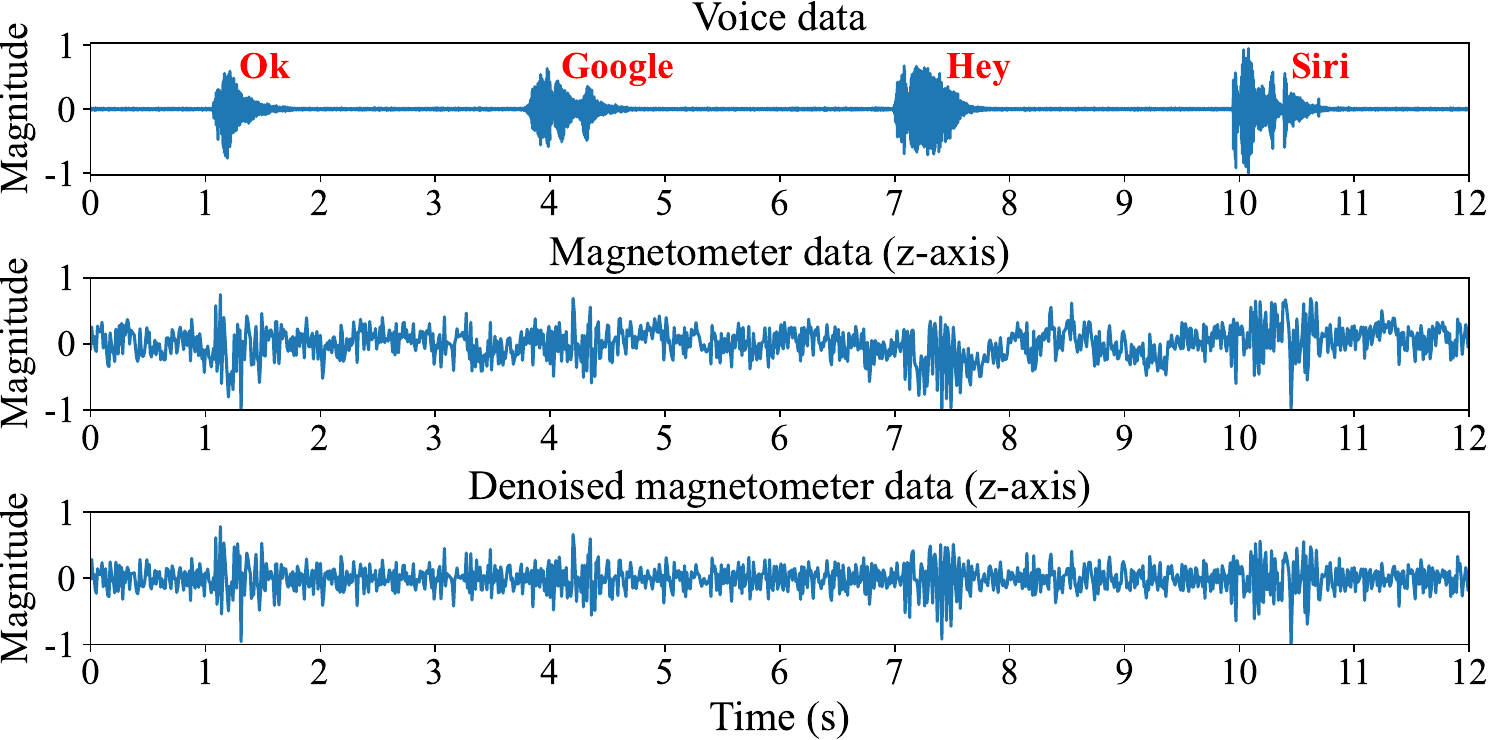}
	\caption{An example of the denoising effect of the z-axis magnetometer data corresponding to the voice command ``OK Google Hey Siri".}
	\label{fig:denoise}
\end{figure}

\section{Design of MagLive}
\label{sec:design}

MagLive is composed of four modules: data capture, data preprocessing, feature extraction, and authentication. 
In this section, we provide a detailed explanation of each module.

\subsection{Data Capture}
MagLive simultaneously collects both voice data and magnetometer data from the authentication device (i.e., smartphone). 
The voice data is utilized for voice authentication, while the magnetometer data is employed for voice liveness detection. 
To account for the intrinsic attenuation properties of the magnetic signal~\cite{liao2022magear}, we conduct sound source distance detection to measure the distance between the sound source and the authentication smartphone, ensuring optimal performance.

\begin{figure}[!t]
	\centering
	\subfloat[Spectrogram of original data]{\includegraphics[width = 0.48\linewidth]{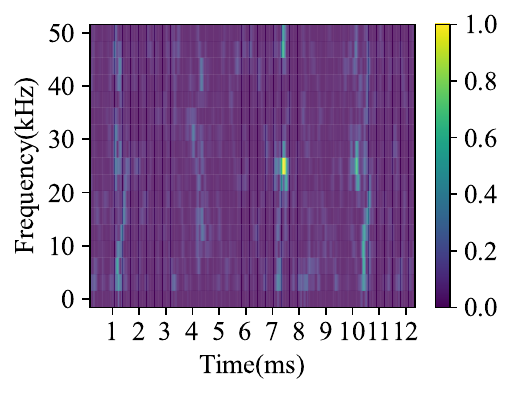}
		\label{fig:before}}
	%	\hspace{0.001em}
	\subfloat[Spectrogram of denoised data]{\includegraphics[width = 0.48\linewidth]{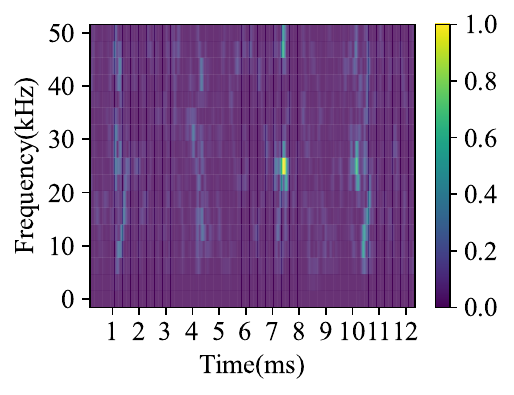}
		\label{fig:after}}
	\caption{An example of the denoising effect applied to the spectrogram of the z-axis magnetometer data.}
	\label{fig:denoise_spectrogram}
\end{figure}

To minimize operational constraints and reduce the need for user cooperation in sound source distance detection, we leverage time delay and energy difference estimation between the two microphones found on commercial smartphones~\cite{wu2022echohand}. 
Our approach eliminates the need for actively transmitting high-frequency acoustic signals and additional user movement, which are required in previous methods~\cite{chen2017you}.
Specifically, we employ generalized cross correlation with phase transformation techniques (GCC-PHAT)~\cite{yang2022deepear} between the signals of two microphones to estimate time difference of arrival (TDOA).
The speed of sound is around 340 m/s at 20$^{\circ}$C~\cite{cai2021active}, allowing us to calculate the distance difference.
If we denote the distances from the sound source to the two microphones as $d_{1}$ and $d_{2}$ respectively, the distance difference can be expressed as $\triangle d = |d_{1} - d_{2}|$.
We denote the energy received by the $i$-th microphone as $E_{i}$.
The relationship between the energies and distances~\cite{tao2022sound} can be obtained as Eq.~\ref{eq:attenuation}:
\begin{equation}
	\label{eq:attenuation}
	E_{1}d_{1}^{2} = E_{2}d_{2}^{2}
\end{equation}
Assuming $d_{1} > d_{2}$, then we have $E_{1} < E_{2}$.
The distances $d_{1}$ and $d_{2}$ can be computed based on Eq.~\ref{eq:distance}:
\begin{equation}
	\label{eq:distance}
	d_{1} = \frac{\sqrt{E_{2}}}{\sqrt{E_{2}}-\sqrt{E_{1}}} \triangle d 
	\quad \textrm{and} \quad
	d_{2} = \frac{\sqrt{E_{1}}}{\sqrt{E_{2}}-\sqrt{E_{1}}} \triangle d
\end{equation}

For more accurate results, it is advisable to maximize the distance difference $\triangle d$, meaning the sound source and the two microphones should ideally be positioned in a straight line.
Our goal is to ensure that the sound source distance remains within a predefined threshold, which is established through experiments as detailed in Section~\ref{sec:robustness}.

\begin{figure}[!t]
	\centering
	\includegraphics[width = 1\linewidth]{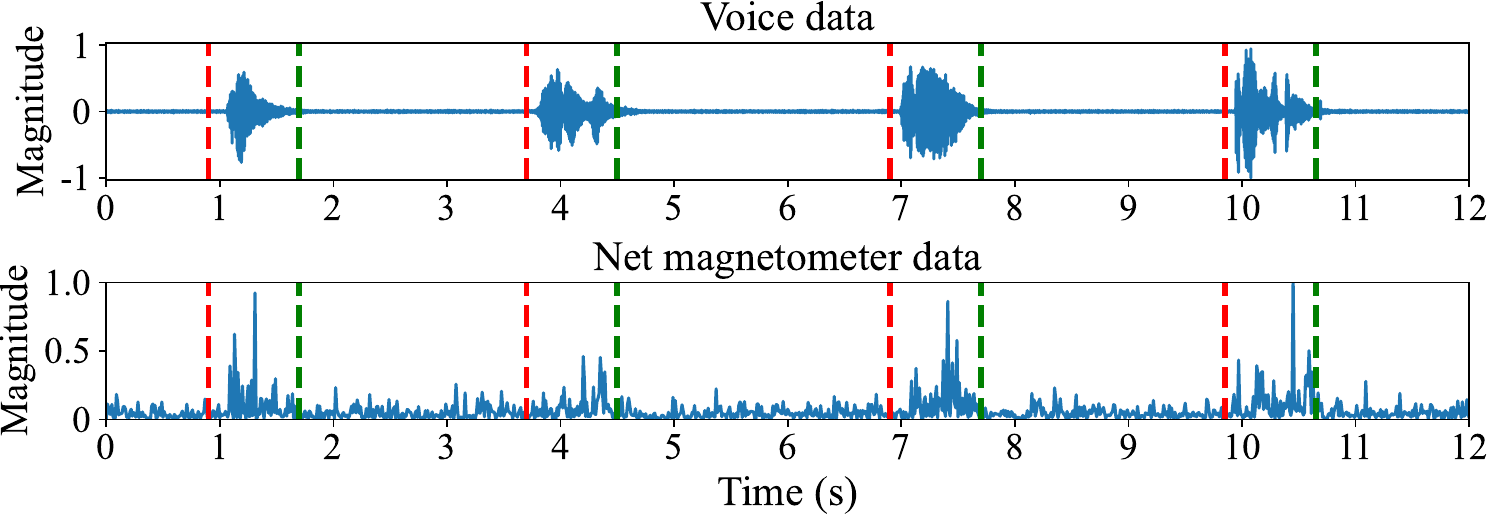}
	\caption{An example of the segmentation result of the net magnetometer data for the voice command ``OK Google Hey Siri", assisted by voice data.}
	\label{fig:segment}
\end{figure}

\subsection{Data Preprocessing}
\label{sec:preprecess}

After capturing the voice and magnetometer data, we perform denoising and use the voice data to assist in segmenting the magnetometer data.

\textbf{Denoising.}
Magnetic field changes induced by speakers are weak and can be easily submerged by noise in the magnetometer~\cite{wang2022automatic}. 
To address this, we first apply a high-pass Butterworth filter~\cite{wang2023voicelistener} to eliminate the noise. 
We set the filter's cut-off frequency to 5 Hz, based on empirical observations and our insights.
Taking the voice command "OK Google Hey Siri" as an example, Fig.~\ref{fig:denoise} demonstrates the denoising effect on the z-axis magnetometer data, while Fig.~\ref{fig:denoise_spectrogram} shows the corresponding effect on the spectrogram. 
We can observe that the denoised magnetometer data exhibits more distinct magnetic patterns related to the voice data.

Considering the influence of the Earth’s magnetic field, the captured three-axis magnetometer data is geo-spatial dependent.
To mitigate the impact of location, we aggregate the magnetometer data across the three axes.
The net magnetometer data $\bm{m^{\prime}}$ derived from the three-axis magnetometer data $\bm{m} = (\bm{m_{x}}, \bm{m_{y}}, \bm{m_{z}}) $ is calculated as follows~\cite{pan2022magdefender}:
\begin{equation}
	\label{eq:total_intensity}
	m^{\prime}(t) = ||\bm{m} (t)|| = \sqrt{m_{x}(t)^2 + m_{y}(t)^2 + m_{z}(t)^2}
\end{equation}

\textbf{Segmentation.}
To accurately pinpoint the magnetic changes induced by speech, we use the voice data to assist in segmenting the magnetometer data. 
First, we apply voice activity detection~\cite{yan2019catcher} to identify segments of speech in the recorded voice data, with each segment indicating the presence of a speech signal.
Since the magnetometer and voice data are synchronized, we then use the voice segments to determine the corresponding segments in the magnetometer data. 
To ensure that each segment of magnetometer data covers an entire speech segment, we adjust the start and end points of each segment to span 100 sample points.
This length is an empirical parameter based on our observations. 
This process ultimately yields magnetometer data segments corresponding to each word for feature extraction.
Fig.~\ref{fig:segment} illustrates an example of the segmentation result of the net magnetometer data, demonstrating the effectiveness of our method.

\begin{figure}[!t]
	\centering
	\includegraphics[width = 0.95\linewidth]{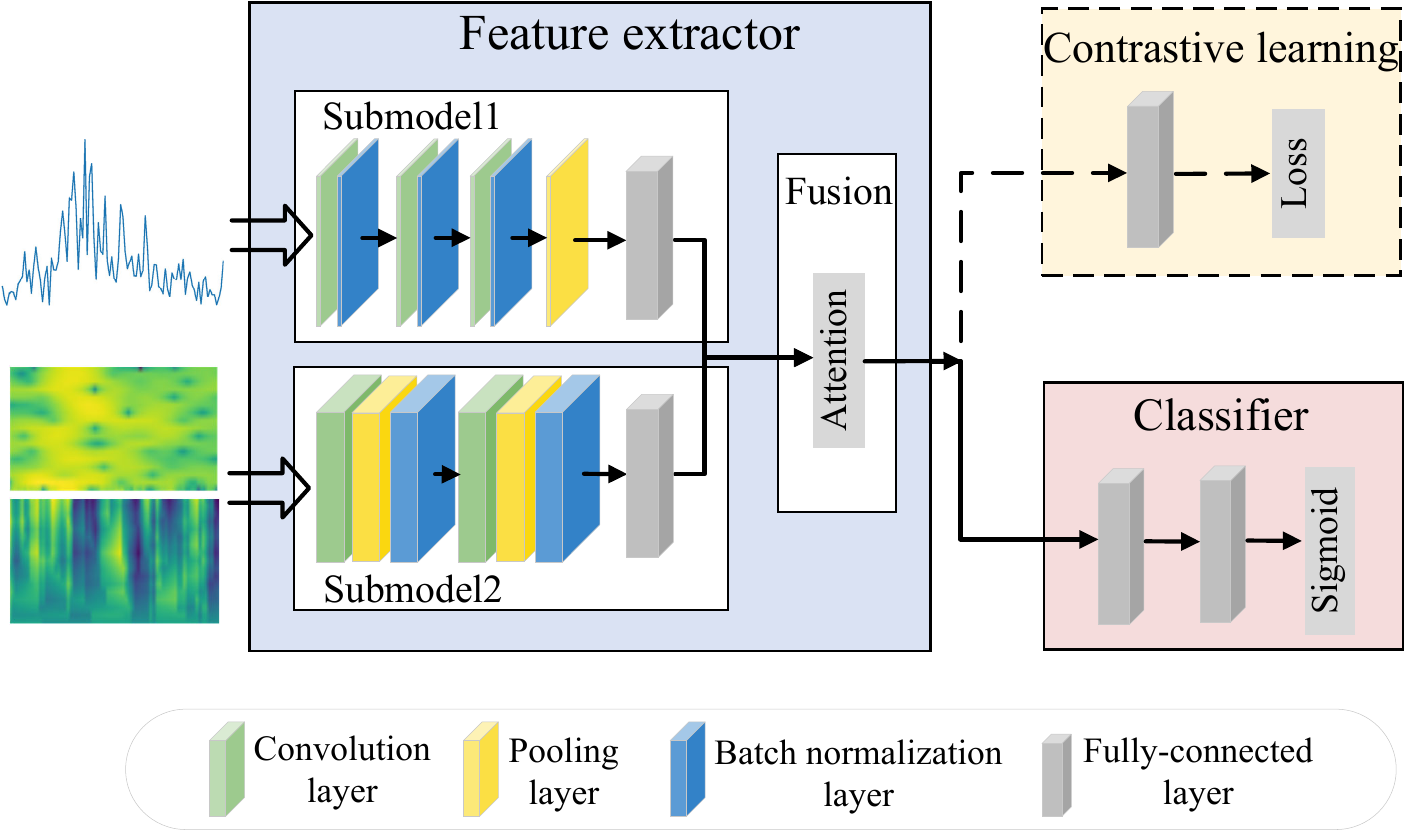}
	\caption{The architecture of our feature extraction and classifier model.}
	\label{fig:structure}
\end{figure}

\begin{table}[!t]
	\centering
	\caption{The structure of our TF-CNN model.}
	\label{tab:cnn}
	\setlength{\tabcolsep}{1.4mm}{
		\begin{tabular}{c|lccr}
			\hline
			Model name & Layer & Layer type  & Output shape & \# Param \\
			\hline
			\multirow{5}{*}{Submodel1} 
			& 1 & Conv1D + BN + ReLU  & (98,16)    & 128    \\
			& 2 & Conv1D + BN + ReLU  & (96,32)    & 1,696  \\
			& 3 & Conv1D + BN + ReLU  & (94,16)    & 1,616  \\
			& 4 & Pooling             & (47,16)    & 0      \\
			& 5 & Flatten + FC + ReLU & (64)       & 48,192 \\
			\hline
			\multirow{5}{*}{Submodel2} 
			& 1 & Conv2D + ReLU       & (15,67,16) & 304    \\
			& 2 & Pooling + BN        & (7,33,16)  & 64     \\
			& 3 & Conv2D + ReLU       & (5,31,32)  & 4,640  \\
			& 4 & Pooling + BN        & (2,15,32)  & 128    \\
			& 5 & Flatten + FC + ReLU & (64)       & 61,504 \\
			\hline
		\end{tabular}
	}
\end{table}

\subsection{Feature Extraction}
We design the TF-CNN-SAF model as a feature extractor to capture magnetic pattern changes generated by different speakers, including both humans and loudspeakers.
Specifically, we first develop a time-frequency convolutional neural network (TF-CNN) to extract feature representations in the time and frequency domains.
Next, we introduce a self-attention-based fusion (SAF) model to learn adaptive weights for feature fusion. 
To further enhance the system's robustness against variations in users, devices, and voice content, we employ a supervised contrastive learning approach, ensuring consistent performance across diverse environments.
Below, we provide a detailed description of our model designs.

\textbf{TF-CNN-based feature representation.}
Deep learning approaches have achieved significant success due to their superior capabilities in feature extraction and representation~\cite{wu2024rethinking,wu2024wafbooster,han2024effectiveness,liang2024towards,liang2024ponziguard,lin2024efficient}.
We design a time-frequency convolutional neural network (TF-CNN) to extract effective magnetic features. 
Specifically, we first extract the envelope from the pre-processed magnetometer data.
Then, we use Short-Time Fourier Transform (STFT)~\cite{wu2020liveness} to generate time-frequency spectrograms for both magnitude and phase information. 
After normalizing the data, we use these as inputs for the feature extraction model.
Since data in the temporal and frequency domains have different dimensions and physical meanings, we design the TF-CNN model with two separate CNN submodels, similar to previous work~\cite{cao2023can}. 
As shown in Fig.~\ref{fig:structure}, the input to submodel1 (i.e., 1D-CNN) is the envelope of the pre-processed magnetometer data, while the input to submodel2 (i.e., 2D-CNN) consists of the time-frequency spectrograms for magnitude and phase.

Table~\ref{tab:cnn} shows the structure of our TF-CNN model.
For submodel1, we first adopt three convolution blocks to learn the feature embedding.
Each convolution block comprises a 1-D convolution (Conv1D) layer, followed by a batch normalization (BN) layer to accelerate the training process, and a ReLU layer as the activation function. 
Subsequently, an average pooling layer is added.
For submodel2, we take 2D spectrograms as input and process them with two 2-D convolution (Conv2D) layers.
For each Conv2D layer, we attach a max pooling layer for dimension reduction and a batch normalization layer to remove the mean and scale the features to unit variance. 
Specifically, the kernel sizes for the Conv2D and max pooling layers are set to $3\times3$ and $2\times2$, respectively.
The feature maps from the two submodels are then flattened and compressed using two fully-connected (FC) layers, resulting in a 128-dimensional concatenated output.

\begin{figure}[!t]
	\centering
	\includegraphics[width = 0.9\linewidth]{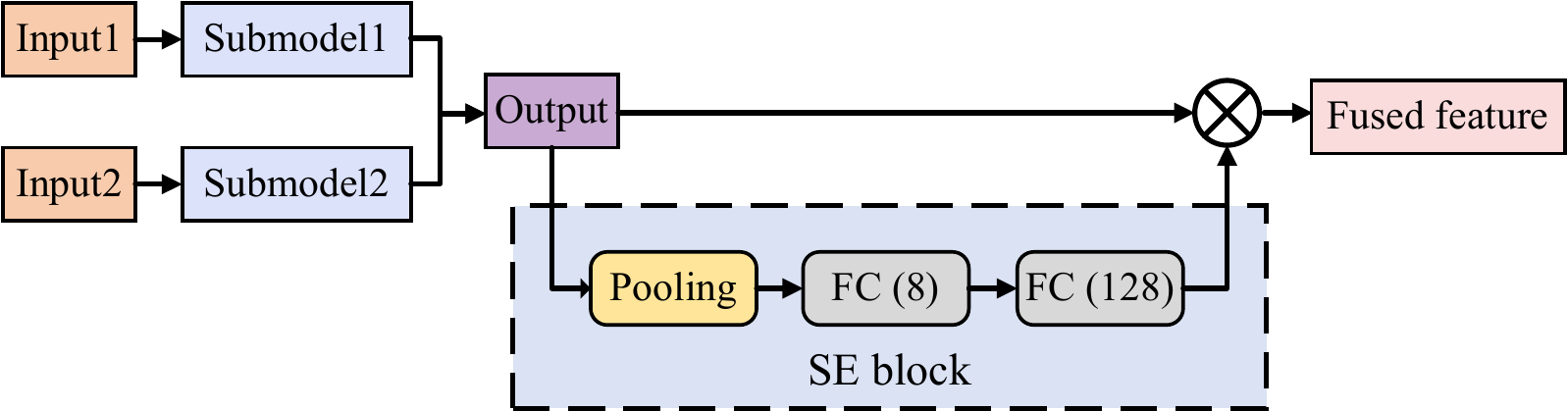}
	\caption{The architecture of our self-attention-based fusion (SAF) model.}
	\label{fig:fuse}
\end{figure}

\textbf{Self-attention-based feature fusion.}
Instead of simply concatenating the features extracted from the two submodels, we develop a self-attention-based fusion (SAF) model, which recalibrates the magnetic features with adaptive weights to selectively emphasize the significant ones.
As shown in Fig.~\ref{fig:fuse}, we employ a Squeeze-and-Excitation (SE) block~\cite{hu2018squeeze} with a global average pooling layer and two fully-connected (FC) layers to learn a weight vector in range of [0, 1].
Specifically, the global average pooling layer computes the average value for each channel, and then two fully-connected (FC) layers produce a weight vector.
The obtained weight vector is used to scale the outputs of the two submodels, creating fused features that assign higher weights to the more informative ones. 
Finally, the feature extractor of MagLive produces a 128-dimensional vector as the feature representation.

\textbf{Contrastive learning-based model training.}
In this paper, our focus lies in detecting the differences in magnetic pattern changes caused by humans and loudspeakers for voice liveness detection.
To construct a user-irrelevant, device-irrelevant, and content-irrelevant liveness detection method, we use supervised contrastive learning~\cite{khosla2020supervised} to train the feature extractor. 
This method eliminates user specificity, device specificity, and content specificity by maximizing the differences between samples from humans and loudspeakers, while minimizing the differences between samples within the same class.

\begin{figure}[!t]
	\centering
	\includegraphics[width = 0.7\linewidth]{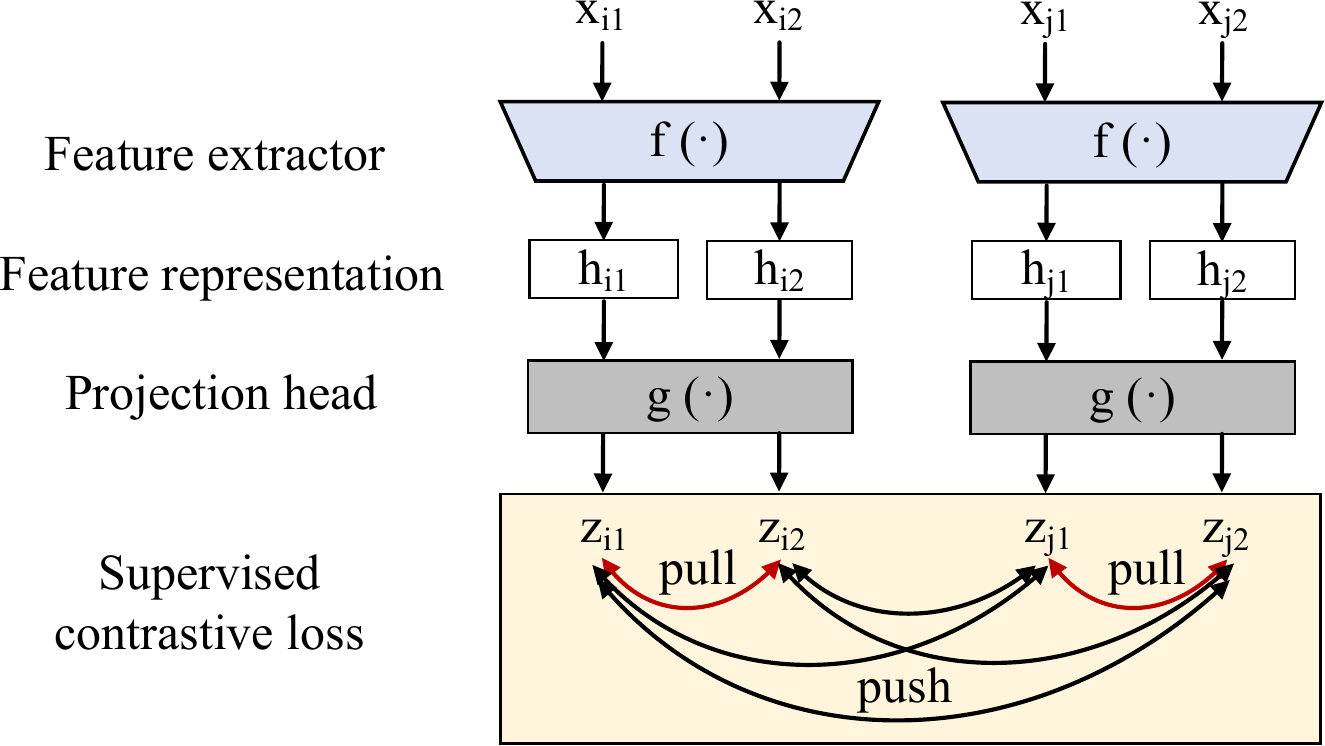}
	\caption{Illustration of contrastive learning-based model training.}
	\label{fig:contrastive}
\end{figure}

\begin{figure}[!t]
	\centering
	\includegraphics[width = 0.65\linewidth]{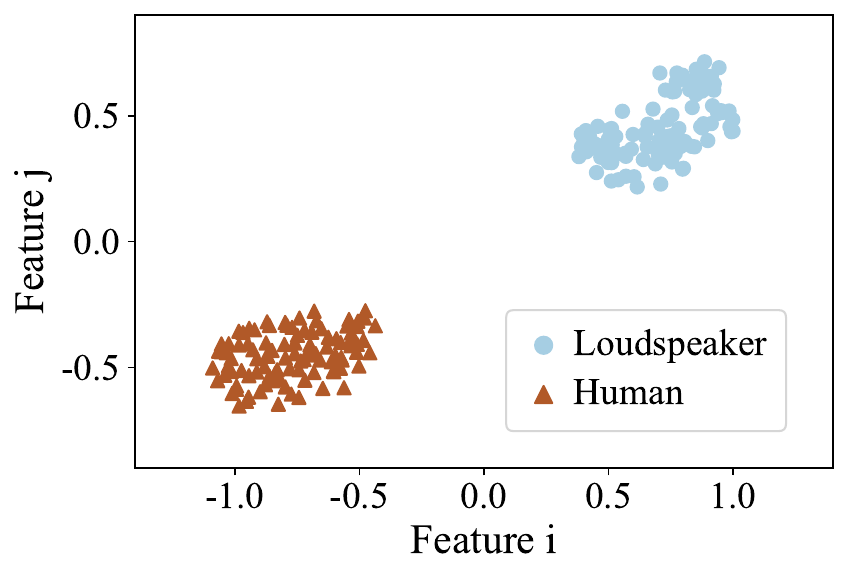}
	\caption{t-SNE visualization of features for the loudspeaker and human.}
	\label{fig:tsne}
\end{figure}

The structure of the contrastive learning network mainly consists of three components: the feature extractor, the projection head, and the supervised contrastive loss.
The feature extractor is built upon the TF-CNN-SAF model.
To compute supervised contrastive loss during training, we append a projection head for the feature extractor, which includes a fully-connected layer and a ReLU layer.
Specifically, for each batch $I$ containing samples and corresponding labels $\{x_i, y_i\}$, where $i \in I \equiv \{1, 2, \ldots, N \}$, the input $x_i$ is first passed through the feature extractor $f(\cdot)$ to obtain the feature representation $h_i$.
Subsequently, the feature representation is further propagated through a projection head $g(\cdot)$, yielding the output $z_i$.
The supervised contrastive loss is then computed on $z_i$ using Eq.~\ref{eq:contrastive_loss}, aiming to minimize the distances between feature representations of samples from the same class while maximizing the distances between those from different classes:
\begin{equation}
	\label{eq:contrastive_loss}
	\mathcal{L} = \sum_{i \in I} \frac{-1}{|P(i)|} \sum_{p \in P(i)} \log \frac{\exp(z_i \cdot z_p/\tau)}{\sum_{a \in A(i)}\exp(z_i \cdot z_a/\tau)}
\end{equation}
Here, $A(i) \equiv I \setminus \{i\}$, $P(i) \equiv \{p \in A(i) : y_p = y_i\}$, and $|P(i)|$ is its cardinality.
The symbol $\cdot$ denotes the inner product, and $\tau$ is a scalar temperature parameter.
In the case where $x_{i1}$ and $x_{i2}$ are from the same class, and $x_{j1}$ and $x_{j2}$ are from another class, as illustrated in Fig.~\ref{fig:contrastive}, the aim is to pull together samples from the same class while pushing apart samples from different classes.

To test the effectiveness of feature representations, we randomly select 200 testing samples from a loudspeaker and a human.
These samples are then fed into the trained feature extractor to extract their corresponding feature representations. 
Then we utilize t-Distributed Stochastic Neighbor Embedding (t-SNE)~\cite{cao2022handkey} to reduce the dimension of the feature representation from 128 to 2 and visualize these samples.
As shown in Fig.~\ref{fig:tsne}, we can see that samples with the same label are closely clustered in the feature space, while samples from different classes are farther apart.
The distinct feature distribution of the loudspeaker and human samples demonstrate the feasibility of our feature extraction approach.

\subsection{Authentication}
Based on the feature representations generated by the feature extractor, MagLive uses binary cross-entropy as the loss function to train a classifier for voice liveness detection, distinguishing between human (1) and loudspeaker (0) samples.
As shown in Fig.~\ref{fig:structure}, the classifier consists of two fully-connected layers followed by a sigmoid layer.
Since voice liveness detection is a binary classification task, the output of the sigmoid layer represents the probability that the voice sample belongs to a human. 
If the probability exceeds a predefined threshold, the sample is classified as originating from a real human. 
Otherwise, it is deemed to be from an attacker (i.e., a loudspeaker). 
In this work, we empirically set the threshold at 0.5.

\begin{table}[!t]
	\centering
	\caption{List of 15 voice commands used in the experiments.}
	\label{tab:command}
	\begin{tabular}{l|l}
		\hline
		No. & Voice commands  \\
		\hline
		1 & Alexa. \\
		2 & Cortana. \\
		3 & OK Google. \\
		4 & Hey Siri. \\
		5 & Hi Assistant. \\
		6 & Turn on Bluetooth. \\
		7 & Take a photo. \\
		8 & Open music player. \\
		9 & Mute the volume. \\
		10 & Show me my messages. \\
		11 & Where is my package? \\
		12 & Call the nearest computer shop. \\
		13 & Remind me to buy milk. \\
		14 & What is my schedule for tomorrow? \\
		15 & What is the time at home? \\
		\hline
	\end{tabular}
\end{table}

\section{Evaluation}
\label{sec:evaluation}

In this section, we report the evaluation results of our proposed voice liveness detection system, MagLive. 
We first introduce the experiment setup, including the data collection and evaluation metrics.
Then we present the overall performance of MagLive and evaluate the security of MagLive in defending against various attacks.
We also conduct comprehensive experiments to evaluate the robustness and effectiveness of MagLive under different settings and factors.

\subsection{Experiment Setup}
\textbf{Data collection.}
Our data collection procedure consists of two phases: human data collection and spoofing data collection.
We use an iPhone 14 Pro as the main authentication device and utilize the Sensor Logger app~\cite{sensorlogger} to gather sensor measurements from the smartphone's built-in sensors.
The built-in magnetometer samples at a rate of 100 Hz, while audio is recorded at 44.1 kHz.

For human data collection, we recruited 20 participants in this study, aged from 21 to 28, including 9 males and 11 females.
We explicitly informed the participants that the purpose of the experiments was to enhance the security of voice authentication on smartphones.
Specifically, our human dataset consists of two parts.
Referring to the WeChat voiceprint~\cite{wechat}, where users read a set of digits as passwords for authentication, each participant was initially requested to say ten digits from zero to nine for 20 times.
Then we selected 15 common voice commands as listed in Table~\ref{tab:command}, and each participant repeated them twice.
Finally, 50 voice commands and corresponding magnetometer data were collected from each participant, resulting in a total of 1000 voice commands (equivalent to 6000 words) and corresponding magnetometer data for the human dataset.
The human dataset was collected in an office room with background noises, including people talking and HVAC noises.

\begin{table}
	\centering
	\caption{Experimental devices and their basic information.}
	\label{tab:device}
	\setlength{\tabcolsep}{1.5mm}{
		\begin{tabular}{l|llll}
			\hline
			No. & Type & Manuf. & Model & Size(L*W*H in cm) \\
			\hline
			1 & Smartphone & Apple & iPhone 14 Pro & 14.8 $\times$ 7.2 $\times$ 0.8 \\
			%		\hline
			2 & Smartphone & Apple & iPhone XR & 15.1 $\times$ 7.6 $\times$ 0.8 \\
			%		\hline
			3 & Smartphone & Huawei & P30 & 14.9 $\times$ 7.1 $\times$ 0.8 \\
			%		\hline
			4 & Smartphone & Google & Pixel 3a & 15.1 $\times$ 7.0 $\times$ 0.8 \\
			%		\hline
			5 & Smartphone & Samsung & Galaxy S10 & 15.0 $\times$ 7.0 $\times$ 0.8 \\
			%		\hline
			6 & Tablet & Apple & iPad Pro & 24.8 $\times$ 17.9 $\times$ 0.6 \\
			%		\hline
			7 & Laptop & Lenovo & ThinkPad X1 & 32.4 $\times$ 21.7 $\times$ 1.6 \\
			%		\hline
			8 & Loudspeaker & Xiaomi & AI Speaker & 8.8 $\times$ 8.8 $\times$ 21.2 \\
			%		\hline
			9 & Loudspeaker & Amazon & Echo Dot & 9.9 $\times$ 9.9 $\times$ 4.3 \\
			\hline
		\end{tabular}
	}
\end{table}

For spoofing data collection, we replayed all the above-collected human voice for each spoofing device. 
Specifically, we employed 8 different spoofing devices varying in size and quality, and the detailed parameters are shown in Table~\ref{tab:device}. 
Since our method is based on the inherent features of loudspeakers, we do not differentiate between spoofing attacks (i.e., replay, speech synthesis, and voice conversion attacks) as they all involve the use of loudspeakers.
In total, we collected 8000 spoofing commands (equivalent to 48000 words) and corresponding magnetometer data.

When evaluating the impacts of various factors on MagLive, we collect additional data, and details are described in the corresponding sections.
Our experiments have received approval from the institutional review board (IRB) of our university.

\textbf{Evaluation metrics.}
We use the following metrics for evaluation.
Balanced accuracy (BAC) measures the overall probability that the system accepts legitimate samples and rejects attack samples. 
It is particularly useful for evaluating the accuracy of imbalanced datasets.
False Acceptance Rate (FAR) represents the rate at which attack samples are wrongly accepted and considered as human samples.
False Rejection Rate (FRR) represents the rate at which legitimate samples are wrongly rejected.
Equal Error Rate (EER)~\cite{wu2020caiauth} shows a balanced view of FAR and FRR, where FAR is equal to FRR.
A satisfactory voice liveness detection method should maintain a high BAC while keeping FAR and FRR low.
The Receiver operation characteristic (ROC) shows the relationship between True Acceptance Rate (TAR, i.e., the proportion of legitimate samples correctly accepted) and the False Acceptance Rate (FAR) at various thresholds~\cite{wu2024s}.
The area under the ROC curve (AUC) is used to measure the probability that prediction scores of legitimate samples are higher than attack samples.

\subsection{Overall Performance}
In this subsection, we evaluate the overall performance of our voice liveness detection system in detecting various attacks.
Specifically, we assess how effectively MagLive can differentiate between authentic human speakers and attackers (i.e., spoofing devices).

\begin{figure}[!t]
	\centering
	\includegraphics[width = 0.7\linewidth]{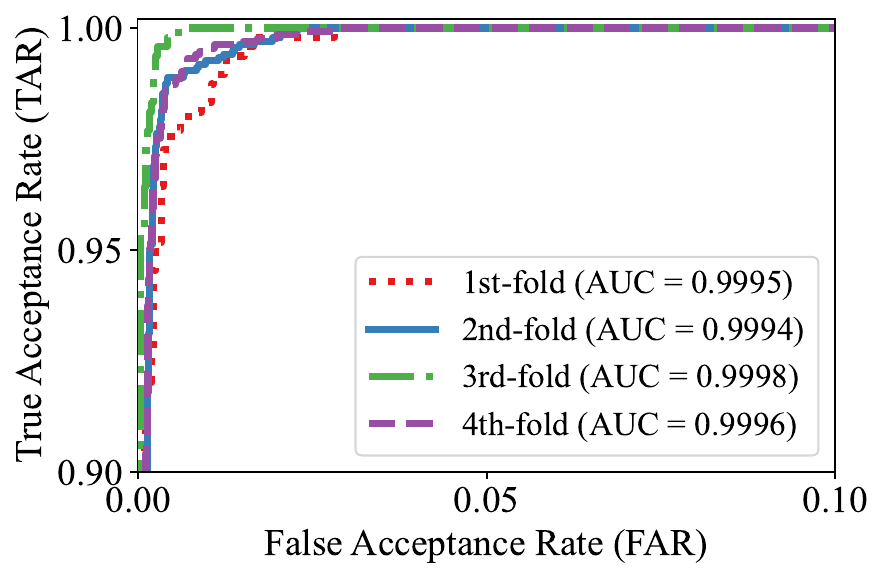}
	\caption{ROC curves of the feature extractor under a 4-fold cross-validation.}
	\label{fig:roc_total}
\end{figure}

\textbf{Performance of feature extractor.}
We utilize a 4-fold cross-validation to evaluate the performance of the feature extractor. 
Specifically, the 20 participants are divided into four groups, each containing five users.
In each fold, we use the data of 5 users to train the feature extractor and then test it with the remaining 15 users.
This process is repeated four times.
Fig.~\ref{fig:roc_total} shows the ROC curves of the feature extractor under the 4-fold cross-validation.
The AUCs are 0.9995, 0.9994, 0.9998, and 0.9996, respectively.
A higher AUC value indicates better system performance.
Table~\ref{tab:feature_extractor} presents the BAC, FAR, FRR, and EER metrics for the 4-fold performance.
MagLive achieves an average BAC of 99.01\% and an EER of 0.77\%.
We use data from the 5 users in the 3rd-fold to train the feature extractor for later evaluation.
Despite being trained on limited data, the feature extractor demonstrates effectiveness to a wide range of users.

\begin{table}[!t]
	\centering
	\caption{BAC (\%), FAR (\%), FRR (\%) and EER (\%) for the feature extractor under a 4-fold cross-validation.}
	\label{tab:feature_extractor}
	\begin{tabular}{l|ccccc}
		\hline
		Metrics & 1st-fold & 2nd-fold & 3rd-fold & 4th-fold & Average \\
		\hline
		BAC (\%) &98.41 &99.05 &99.55 &99.02 &\textbf{99.01} \\
		FAR (\%) &0.44 &0.35 &0.28 &0.38 &\textbf{0.36} \\
		FRR (\%) &2.73 &1.55 &0.63 &1.58 &\textbf{1.62} \\
		EER (\%) &1.08 &0.87 &0.41 &0.71 &\textbf{0.77} \\
		\hline
	\end{tabular}
\end{table}

\begin{figure}[!t]
	\centering
	\includegraphics[width=1\linewidth]{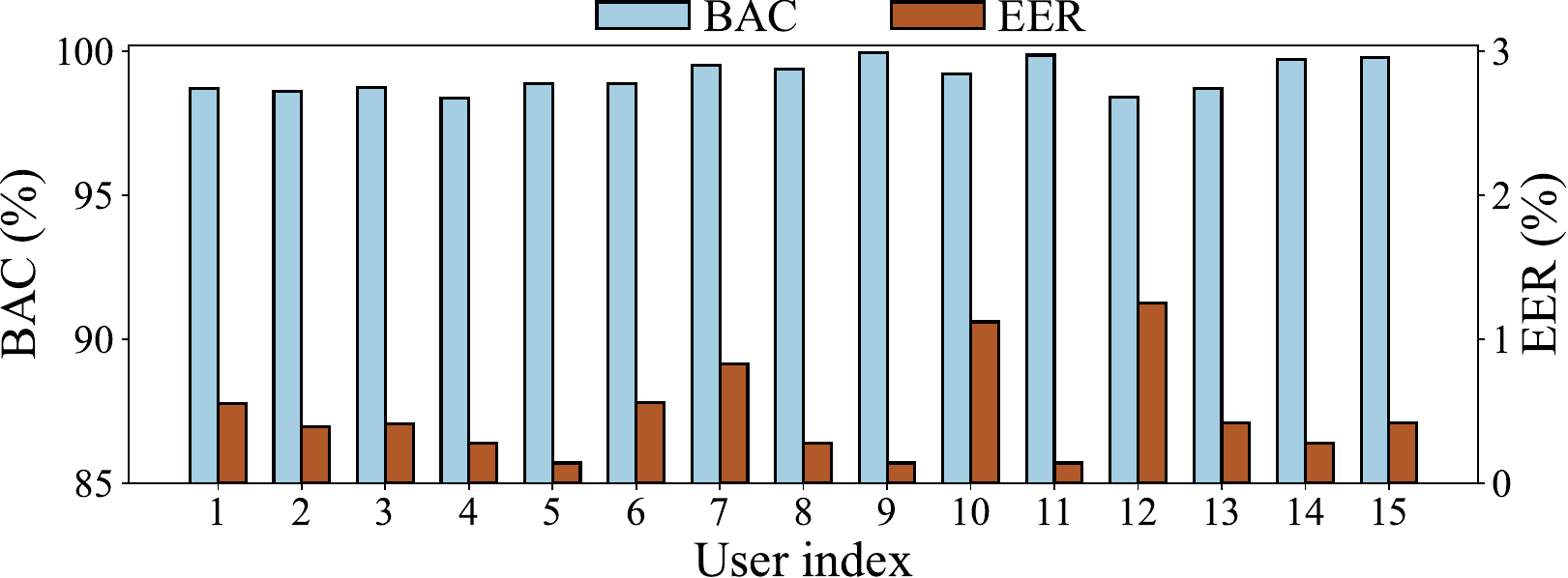}
	\caption{The BAC and EER performance for each user.}
	\label{fig:per_user}
\end{figure}

\textbf{Per-user breakdown analysis.}
To evaluate the performance of MagLive on 15 different users, we train a classifer for each user using their own legitimate data and spoofing data. 
Fig.~\ref{fig:per_user} shows the BAC and EER of each user. 
As we can see, the best case among the 15 users is user 9, achieving a remarkable BAC of 99.93\% and an impressively low EER of 0.14\%.
Even in the worst case (i.e., user 12), the BAC remains over 98.3\% with an EER of less than 1.3\%.
Despite varying performance across users, the average BAC of 99.11\% and EER of 0.48\%, demonstrate MagLive's effectiveness in distinguishing humans from loudspeakers.

\begin{figure*}
	\begin{minipage}[t]{0.32\linewidth}
		\centering
		\includegraphics[width = 1\linewidth]{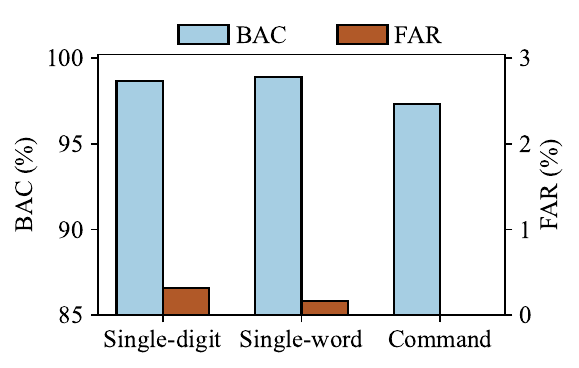}
		\caption{The BAC and FAR performance for different voice content.}
		\label{fig:voice_content}
	\end{minipage}
	\hspace{0.5em}
	\begin{minipage}[t]{0.32\linewidth}
		\centering
		\includegraphics[width = 1\linewidth]{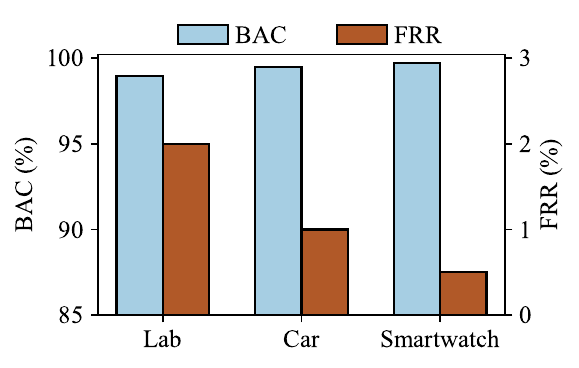}
		\caption{The BAC and FRR performance for different environments.}
		\label{fig:mag_env}
	\end{minipage}
	\hspace{0.5em}
	\begin{minipage}[t]{0.32\linewidth}
		\centering
		\includegraphics[width = 1\linewidth]{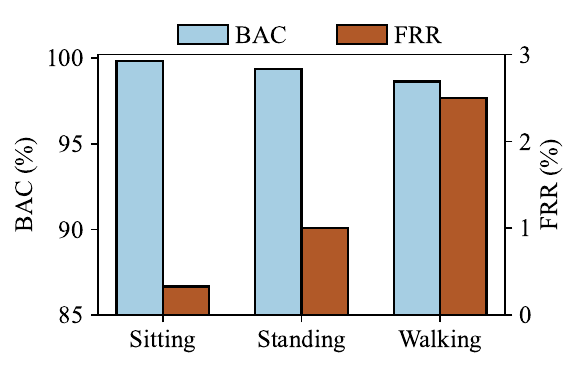}
		\caption{The BAC and FRR performance under different postures.}
		\label{fig:posture}
	\end{minipage}
\end{figure*}

\textbf{Performance for different voice content.}
Just like Voshield~\cite{yang2023voshield}, which uses the clip of a voice command for voice liveness detection, we first evaluate the performance single-digit and single-word performance for MagLive.
Considering data from 15 participants, we train the classifier with different types of voice content, respectively.
Additionally, we evaluate the performance of entire voice commands~\cite{chen2023rf}, as shown in Table~\ref{tab:command}.
We assume that once a word is identified as a spoofing sample, the whole voice command is recognized as a spoofing one.
Fig.~\ref{fig:voice_content} displays the BACs and FARs for different voice content.
For single-digits, single-words, and voice commands, the BACs are 98.67\%, 98.91\%, and 97.33\%, respectively, with FARs of 0.32\%, 0.17\%, and 0\%.
While there's a slight decrease in BAC for the whole voice command, MagLive successfully detects all the spoofing samples.

\textbf{Defending against advanced attacks.}
Besides spoofing attacks, we also evaluate the effectiveness of MagLive on thwarting advanced attacks, such as modulated attacks~\cite{wang2020differences} and adversarial attacks~\cite{yu2023smack}.
Modulated attacks~\cite{wang2020differences} modify the voice spectrum to relieve the distortion induced by loudspeakers.
Adversarial attacks~\cite{yu2023smack} preserve semantic information of human speech to achieve naturalness in the resulted adversarial examples.
In this experiment, we use human data from a participant as legitimate data and select the Huawei P30 as the spoofing device to generate 200 spoofing samples for each type of attack.
The BACs are 99.5\% and 100\%, with FARs of 0\%.
MagLive is effective in detecting advanced attacks because these attacks inevitably involve loudspeakers, leading to magnetic pattern changes during playback.

\begin{figure}[!t]
	\centering
	\includegraphics[width=1\linewidth]{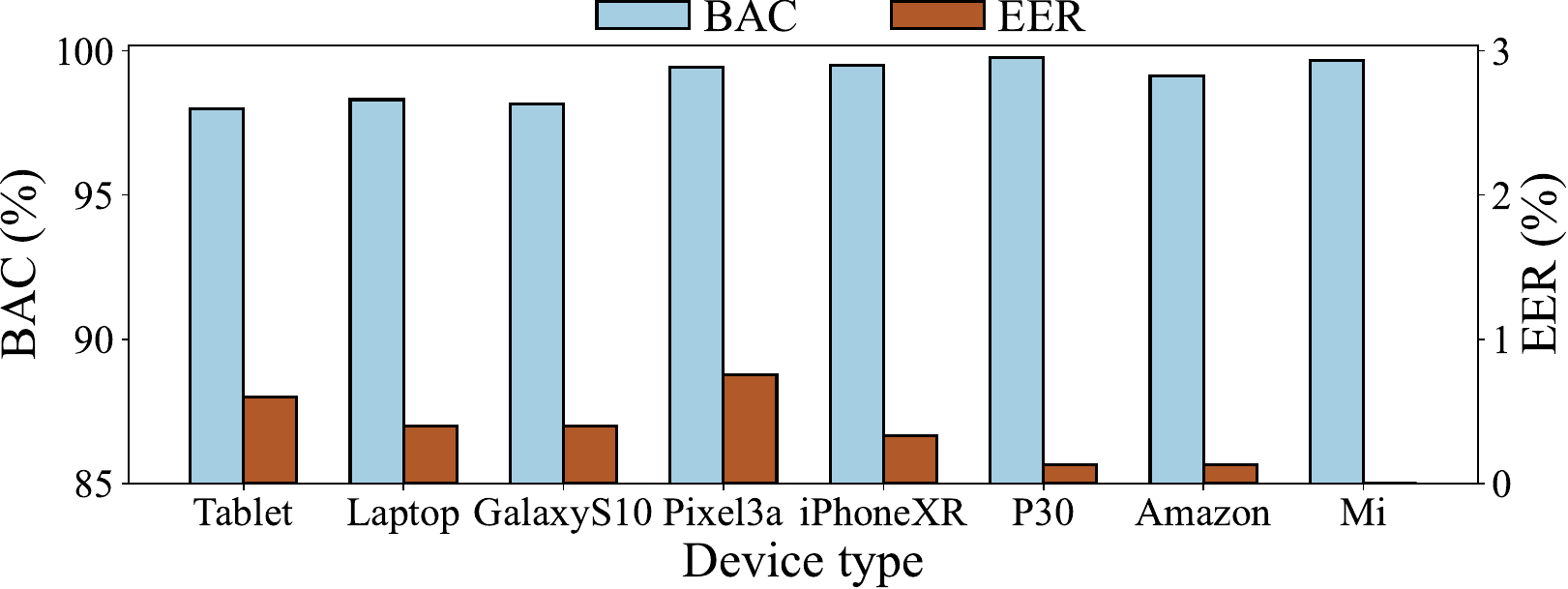}
	\caption{The BAC and EER performance for different spoofing devices.}
	\label{fig:spoofing_devices}
\end{figure}

\subsection{Robustness of MagLive}
\label{sec:robustness}
In this subsection, we evaluate the performance of MagLive under different settings and factors, which demonstrate the robustness and effectiveness of MagLive.

\textbf{Impact of spoofing devices.}
We investigate the liveness detection performance when the attacker uses different types of devices.
Table~\ref{tab:device} lists 8 spoofing devices used in this study, including four smartphones, a tablet, a laptop, and two smart loudspeakers.
These devices represent common types that an attacker might employ.
To evaluate the performance of a spoofing device, we train the classifier using data from the remaining 7 spoofing devices.
Fig.~\ref{fig:spoofing_devices} illustrates the BAC and EER of MagLive for each device in this case. 
Among the 8 devices, smart loudspeakers are more easily detected, possibly due to their stronger magnetic effect. 
However, even the worst-performing device achieves a BAC of over 98\% and an EER of less than 0.75\%.
Overall, MagLive demonstrates robustness to various spoofing devices and achieves device-irrelevance.

\begin{figure}[!t]
	\centering
	\includegraphics[width=1\linewidth]{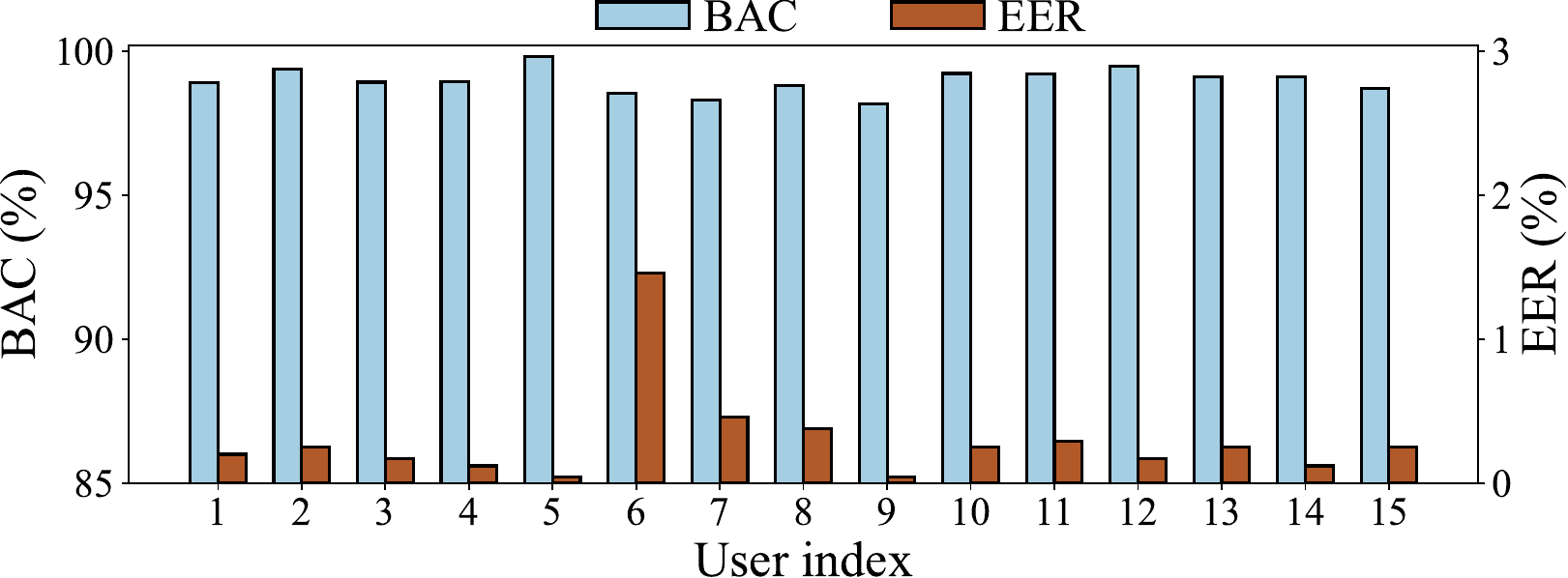}
	\caption{The BAC and EER performance of cross-user training.}
	\label{fig:cross_user}
\end{figure}

\textbf{Impact of cross-user training.}
To evaluate the cross-user performance of MagLive, we train the classifer with legitimate data and spoofing data from 14 users, and then test it with the data from a remaining unseen user.
Fig.~\ref{fig:cross_user} shows the BACs and EERs for different users. 
All users exhibit good performance, achieving a BAC of over 98\% and an EER of less than 1.5\%. 
In summary, MagLive demonstrates high user-irrelevant performance.

\textbf{Impact of speech type.}
Our dataset consists of two types of speech: digits and voice commands.
Initially, we train the classifier using the entire data of digits and test it on the voice command data. 
The resulting BAC is 99.36\%, with an EER of 0.45\%. 
Subsequently, we train the classifier using the voice command data and test it on the digit data. In this case, the BAC is 98.47\%, with an EER of 0.63\%. 
These results demonstrate that MagLive can achieve content-irrelevant performance.

\textbf{Impact of environments.}
To evaluate the impact of environments, we conduct experiments in three different scenarios.
Firstly, our method is evaluated in a laboratory environment surrounded by various electrical devices, such as computers, wireless keyboards, and wireless mice. 
Secondly, we replicate the experiment in a Volkswagen Bora car.
Finally, we examine the scenario where a user wears a smart wristband. 
Fig.~\ref{fig:mag_env} shows the BACs and FRRs for different magnetic field environments. 
The BACs for three environments are 98.96\%, 99.46\%, and 99.71\%, respectively.
All FRRs remain below 2\%. 
These results demonstrate MagLive is available for different magnetic environments.

\textbf{Impact of user posture.}
In the overall performance evaluation, the participants adopt sitting postures.
In this experiment, we also consider the standing and walking postures to evaluate the impact of different user postures.
For this analysis, we assess one participant and 8 spoofing devices.
As shown in Fig.~\ref{fig:posture}, MagLive exhibits better performance in sitting and standing postures compared to walking, suggesting potential interference from user movement. 
To mitigate this fluctuation, we intend to develop a model that learns movement patterns to effectively filter out noise caused by motion.

\begin{figure}
	\begin{minipage}[t]{0.495\linewidth}
		\centering
		\includegraphics[width = 1\linewidth]{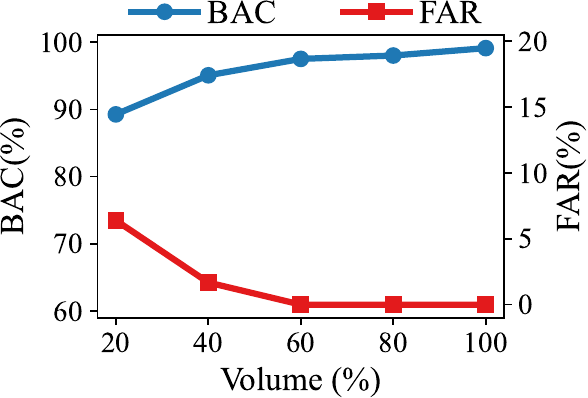}
		\caption{The BAC and FAR at different volume levels.}
		\label{fig:volume}
	\end{minipage}
	\hfill
	\begin{minipage}[t]{0.495\linewidth}
		\centering
		\includegraphics[width = 1\linewidth]{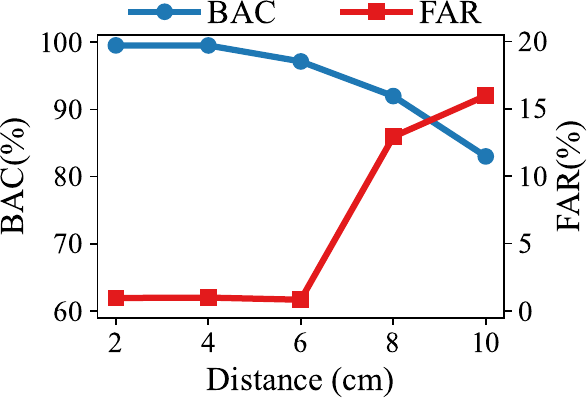}
		\caption{The BAC and FAR at different distances.}
		\label{fig:distance}
	\end{minipage}
\end{figure}

\textbf{Impact of voice volume.}
We evaluate the performance of the smartphone’s magnetometer at different volume levels. 
Specifically, we use Huawei P30 as the spoofing device to replay recordings at 20\%, 40\%, 60\%, 80\%, and 100\% of the maximum volume supported by the smartphone.
As shown in Fig.~\ref{fig:volume}, as the volume increases from 20\% to 100\%, the BAC improves from 89.25\% to 99.1\%, while the FAR decreases from 6.4\% to 0\%.
A higher voice volume could lead to larger magnetic changes.
Typically, human normal conversation exceeds 40 dB~\cite{chen2023rf}, which corresponds to over 40\% of the maximum smartphone volume.
These results present that MagLive is available for various volume levels.

\textbf{Impact of distance threshold.}
We study the impact of the distance between the sound source and the authentication device (i.e. smartphone) to determine an appropriate distance threshold.
We recruit a participant to speak from various distances, and the Huawei P30 replays these voice samples accordingly.
Fig.~\ref{fig:distance} shows the BACs and FARs of MagLive at different authentication distances.
Notably, for better performance (e.g., BAC over 97\%), the results are most satisfactory within 6 cm.
To Balance security and user-friendliness, we ultimately set the threshold at 6 cm for optimal performance.

\textbf{Impact of authentication smartphones.}
To explore the influence of different authentication smartphones, we introduce two additional devices: iPhone XR and Huawei P30. 
In this setup, we involve one participant and utilize the Galaxy S10 as the spoofing device.
The BACs are 96.22\% and 96.5\%, with corresponding EERs of 1.48\% and 1.5\%, respectively.
Both smartphones yield comparable performance to that of the iPhone 14 Pro. 
These results indicate that MagLive is effective across different authentication smartphones.

\begin{table*}[!t]
	\centering
	\caption{Comparison of state-of-the-art voice liveness detection methods}
	\label{tab:related work}
	\setlength{\tabcolsep}{0.9mm}{
		\begin{tabular}{c|llccccccc}
			\hline
			Device & System   & Distinctiveness & \makecell[c]{No active\\sensing} & \makecell[c]{No specialized \\ hardware} & \makecell[c]{Little usage \\ constraint} & \makecell[c]{Resist spectrum \\ modulated attacks} & \makecell[c]{Diverse \\ environments} & Accuracy & EER\\
			\hline
			\multirow{7}{*}{Smartphones} 
			& CaField~\cite{yan2019catcher} & Sound field & \checkmark & \checkmark & $\times$ & \checkmark & \checkmark &99.16\% &0.85\% \\
			& VoiceGesture~\cite{zhang2017hearing} & Mouth motion & $\times$ & \checkmark & \checkmark & \checkmark & \checkmark & $\sim$ 99\% & $\sim$ 1\% \\
			& VoiceLive~\cite{zhang2016voicelive} & Phoneme location & \checkmark & \checkmark & $\times$ & \checkmark & \checkmark & $\sim$ 99\% & $\sim$ 1\% \\
			& Wang \emph{et al.}~\cite{wang2019secure} & Oral airflow & \checkmark & $\times$ & \checkmark & \checkmark & \checkmark & 97.25\% & 2.08\% \\
			& Void~\cite{ahmed2020void} & Hardware imperfections & \checkmark & \checkmark & \checkmark & $\times$ & \checkmark & $\sim$ 98\% & $\sim$ 1\% \\
			& Chen \emph{et al.}~\cite{chen2017you} & Magnetic field & $\times$ & \checkmark & $\times$ & \checkmark & $\times$ & 100\% & 0\% \\
			& MagLive (our work) & Magnetic pattern changes & \checkmark & \checkmark & \checkmark & \checkmark & \checkmark & 99.01\% & 0.77\%\\
			\hline
			\multirow{2}{*}{Other devices} 
			& ArrayID~\cite{meng2022your} & Multi-channel audio  & \checkmark & $\times$ & \checkmark & \checkmark & \checkmark & 99.84\% & 0.17\% \\
			& WearID~\cite{shi2020wearid} & Device ownership & \checkmark & $\times$ & \checkmark & \checkmark & \checkmark & 97.2\% & N/A \\
			\hline
		\end{tabular}
	}
\end{table*}

\subsection{Overhead}
We implemented a prototype of MagLive to evaluate its authentication latency and computational overhead on Huawei P30. 
After capturing the data, we evaluated the processing time for a single data segment and calculated the average latency over 10 trials. 
MagLive achieves an average authentication latency of 0.1 seconds.
Additionally, we used the Android Profiler tool to assess the computational overhead of MagLive. 
The results show that MagLive requires an average memory usage of around 59.4 MB during the authentication process.

\section{Related work}
\label{sec:related Work}

Liveness detection enhances the security of voice authentication against spoofing attacks.
In this section, we categorize existing voice liveness detection methods into two classes based on the authentication device type: smartphones and other devices (e.g., smart speakers and wearable devices).
Table~\ref{tab:related work} summarizes the characteristics of some state-of-the-art voice liveness detection methods.

\subsection{Voice Liveness Detection on Smartphones}
For voice liveness detection methods on smartphones, we further divide them into three classes based on the source of distinctiveness: sound field features~\cite{yan2019catcher}, human features~\cite{zhang2017hearing,lu2018lippass,wu2019lvid,chen2021chestlive,zhang2016voicelive,cao2022liveprobe,wang2019voicepop,wang2019secure,blue2022you} and loudspeaker features~\cite{blue2018hello,ahmed2020void,chen2017you}.

\textbf{Sound field features.}
These methods use the unique biometric information embedded in the sound field during the sound propagation stage.
CaField~\cite{yan2019catcher} extracts the fieldprint from the sound field to detect speakers (either humans or loudspeakers).
However, users are required to maintain a fixed manner to ensure the robustness of these fieldprints.

\textbf{Human features.}
These methods focus on unique features during human voice generation for liveness detection.
Some methods use smartphones to actively transmit high-frequency acoustic signals to sense user’s articulatory gestures~\cite{zhang2017hearing}, lip motions~\cite{lu2018lippass,wu2019lvid} and chest motions~\cite{chen2021chestlive} during human voice generation.
VoiceLive~\cite{zhang2016voicelive} measures the time-difference-of-arrival (TDoA) changes in a sequence of phoneme sounds, which is sensitive to the placement of the smartphone.
Some methods leverage the unique energy responses~\cite{cao2022liveprobe}, breathing pop sounds~\cite{wang2019voicepop}, and the arrangement of the human vocal tract estimated by fluid dynamics~\cite{blue2022you} during speech generation.
Wang \emph{et al.}~\cite{wang2019secure} detect the pressure of the oral airflow using specialized hardware.

\textbf{Loudspeaker features.}
Researchers also detect voice liveness by concentrating on features that arise when a loudspeaker generates sound.
Blue \emph{et al.}~\cite{blue2018hello} and Void~\cite{ahmed2020void} use distortions in the sound caused by the hardware imperfections of loudspeakers as features. 
These methods only rely on features extracted in the audio domain, making them vulnerable to acoustic attacks, such as spectrum modulated attacks~\cite{wang2020differences}.
Most loudspeakers generate magnetic fields when producing sounds.
Chen \emph{et al.}~\cite{chen2017you} use the absolute value and changing rate of the magnetic field to distinguish between humans and loudspeakers.
It requires actively transmitting high-frequency acoustic signals and specific user actions to move the smartphone along a predefined trajectory.
Additionally, it relies on thresholds for magnetic strength and rate of change, which limits its applicability in diverse environments.
In contrast, our work is the first to explore the effective changes in magnetic patterns associated with speech.
MagLive eliminates the need for active sensing and specialized hardware, imposing little usage constraint.
It is robust against various attacks and effective across diverse environmental conditions.

\subsection{Voice Liveness Detection on Other Devices}
Existing methods utilize microphone arrays~\cite{li2021robust,meng2022your,yang2023voshield,lee2020using} and specialized hardware~\cite{meng2018wivo,zhao2021anti,pradhan2019combating,li2020vocalprint} for voice liveness detection on smart speakers.
Additionally, some methods require extra wearable devices to assist authentication~\cite{shi2020wearid,feng2017continuous,blue20182ma}.

Li \emph{et al.}~\cite{li2021robust}, ArrayID~\cite{meng2022your} and VoShield~\cite{yang2023voshield} utilize multi-channel audio from the microphone array to detect the difference between human speech and spoofing audio.
Speaker-Sonar~\cite{lee2020using} leverages a circular microphone array to emit the inaudible sound and track the user’s direction. 
All these methods require microphone arrays for detection.
Some methods capture wireless signals to recognize mouth motions~\cite{meng2018wivo,zhao2021anti} or unique breathing rates~\cite{pradhan2019combating} to distinguish authentic voice commands from spoofed ones.
Vocalprint~\cite{li2020vocalprint} leverages skin-reflect mmWave signals to sense the minute vocal vibrations in the near-throat region of users.
These methods require additional sensors such as wireless sensors and mmWave radars.
WearID~\cite{shi2020wearid} leverages motion sensors on the user's wearable device to capture aerial speech in the vibration domain and verifies it with the speech captured in the audio domain.
VAuth~\cite{feng2017continuous} 
is designed to fit in widely-adopted wearable devices, such as eyeglasses and earphones. 
2MA~\cite{blue20182ma} 
uses multiple devices operating in the same area to eliminate replay attacks.
These methods require additional smart devices to assist with authentication, leading to extra costs.

\section{Discussion}
\label{sec:discussion}

In this section, we discuss some limitations in our work and experiments, and provide an outlook for potential improvements in future work. 
Our study addresses a realistic spoofing attack scenario, where attackers record and manipulate voice samples, then replay them with loudspeakers.
We use the smartphone's built-in magnetometer to capture magnetic pattern changes for voice liveness detection.
For better performance (e.g., BAC over 97\%), we set the distance threshold between the sound source and the authentication smartphone at 6 cm. 
We acknowledge that MagLive does not cover all usage scenarios involving other types of smart devices, such as wearable devices and smart speakers used over remote distances.
Our method for detecting sound source distance is lightweight and operates on a 2D plane. 
To enhance accuracy and universality, future work could explore incorporating deep learning technologies~\cite{lin2024adaptsfl,lin2023fedsn,lin2024split,lin2023pushing,lin2024splitlora,fang2024automated} to develop a more sophisticated detection scheme.

Our segmentation and denoising method is based on empirical observations and insights, and we have not yet evaluated its performance through experiments. 
User movement can potentially interfere with the performance of MagLive. 
We plan to develop a model that learns movement patterns to effectively filter out noise caused by motion.
Previous studies have shown that magnetic fields can never be completely eliminated, and simple magnetic shielding measures have limited effectiveness in evading detection~\cite{liu2023camradar}.
In future work, we aim to conduct additional experiments to explore the impact of magnetic field shielding on MagLive.

Although we have collected datasets from a group of students in our experiment, larger-scale and longer-term experiments are necessary to ensure the applicability of MagLive in real-world scenarios. 
We recognize that a larger sample size would yield more robust and generalizable results. 
In future work, we plan to expand the study to include more participants to further validate our findings.

\section{Conclusion}
\label{sec:conclusion}
In this paper, we propose MagLive, a robust voice liveness detection system on smartphones.
MagLive utilizes the smartphone's built-in magnetometer to capture magnetic pattern changes associated with speech, eliminating the need for active sensing. 
By employing TF-CNN-based deep learning, self-attention-based fusion mechanisms, and supervised contrastive learning, MagLive extracts effective and robust magnetic features, achieving consistent performance in various environments.
MagLive requires no specialized hardware, imposes minimal usage constraints, and demonstrates resilience against attacks. 
Overall, MagLive presents a promising security enhancement for existing voice authentication systems on smartphones.

\bibliographystyle{IEEEtran}
\bibliography{arxiv}

\end{document}